\documentclass{elsart}
\usepackage{graphicx}
\usepackage{amssymb}
\begin{document}

\begin{frontmatter}
\title{Analysis of the low-energy $\pi^- p$ charge-exchange data}
\author[EM]{E. Matsinos{$^*$}},
\author[GR]{G. Rasche},
\address[EM]{Centre for Applied Mathematics and Physics, Zurich University of Applied Sciences, Technikumstrasse 9, P.O. Box, CH-8401 Winterthur, Switzerland}
\address[GR]{Institut f\"{u}r Theoretische Physik der Universit\"{a}t, Winterthurerstrasse 190, CH-8057 Z\"{u}rich, Switzerland}

\begin{abstract}
We analyse the charge-exchange (CX) measurements $\pi^- p\rightarrow \pi^0 n$ below pion laboratory kinetic energy of $100$ MeV. After the removal of five degrees of freedom from the initial database, we combine it with the truncated 
$\pi^+ p$ database of Ref.~\cite{mrw1} and fit the ETH model \cite{glmbg} to the resulting data. The set of the parameter values of the ETH model, as well as the predictions derived on their basis for the hadronic phase shifts and for 
the low-energy $\pi N$ constants, are significantly different from the results obtained in the analysis of the truncated $\pi^\pm p$ elastic-scattering databases. The main difference in the hadronic phase shifts occurs in 
$\tilde{\delta}_{0+}^{1/2}$. We discuss the implications of these findings in terms of the violation of the isospin invariance in the hadronic part of the $\pi N$ interaction. The effect observed amounts to the level of $7-8 \%$ in the 
CX scattering amplitude below $70$ MeV. The results and conclusions of this study agree well with those obtained in the mid 1990s, when the isospin invariance was first tested by using $\pi N$ experimental data, and disagree with the 
predictions obtained within the framework of the heavy-baryon Chiral-Perturbation Theory.
\end{abstract}
{$^*$}{Corresponding author. E-mail: evangelos.matsinos@zhaw.ch, evangelos.matsinos@sunrise.ch; Tel.: +41 58 9347882; Fax: +41 58 9357306}
\end{frontmatter}

\section{\label{sec:Introduction}Introduction}

This is the last of three papers addressing issues of the pion-nucleon ($\pi N$) interaction at low energies (pion laboratory kinetic energy $T \leq 100$ MeV). In the first paper \cite{mrw1}, we reported on a new phase-shift analysis (PSA) 
of the $\pi^\pm p$ elastic-scattering measurements. In the second paper \cite{mrw2}, we examined the self-consistency of the $\pi^\pm p$ elastic-scattering differential cross sections of Ref.~\cite{denz}, which we have not included in our 
PSAs. In the present study, we will analyse the experimental data for the charge-exchange (CX) reaction $\pi^- p \rightarrow \pi^0 n$ and investigate whether earlier claims on the isospin breaking \cite{glk,m} need to be revised in view of 
the impressive increase of the CX database and of the development of our analysis methods during the last fifteen years.

We will mark the physical quantities extracted in the PSA of the $\pi^\pm p$ elastic-scattering data \cite{mrw1} with the label `ZUAS12'; this applies both to the solution obtained for the parameters of the ETH model \cite{glmbg}, as well as 
to the predictions derived on the basis of Tables 3 (for $\mathrm{p}_{min} \approx 1.24 \cdot 10^{-2}$, where $\mathrm{p}_{min}$ denotes the confidence level for the acceptance of the null hypothesis, i.e., of no statistically-significant 
effects) and 4 of Ref.~\cite{mrw1}. The corresponding results, obtained in the present work from the common analysis of the $\pi^+ p$ and CX databases, will be marked with the label `ZUAS12a'.

Similarly to Ref.~\cite{mrw1}, we will assume that the physical quantities appearing in the present study (i.e., the fit parameters of Sections \ref{sec:K-Matrix_CX} and \ref{sec:Model}, the scattering lengths and volumes of Section 
\ref{sec:ModelConstants}, the hadronic phase shifts of Section \ref{sec:ModelPhases}, etc.) are not purely-hadronic quantities since they still contain residual electromagnetic (em) effects. The repetitive use of the term `em-modified' is 
clumsy; therefore, we will omit it, unless we consider its use necessary as, for instance, in the captions of the tables and figures, as well as in Section \ref{sec:Discussion}.

\section{\label{sec:Method}Method}

The formalism which we use here has been described in detail in Ref.~\cite{mworg}. The determination of the observables from the hadronic phase shifts may be found in Section 2 of that work. For $\pi^+ p$ scattering, one obtains the partial-wave 
amplitudes from Eq.~(1) and determines the no-spin-flip and spin-flip amplitudes via Eqs.~(2) and (3). The observables are obtained from these amplitudes via Eqs.~(13) and (14). For the CX reaction, the observables are determined using Eqs.~(21-24) 
of Ref.~\cite{mworg}.

All the details on the analysis method (i.e., on the minimisation function, on the scale factors, etc.) may be found in Section 2.2 of Ref.~\cite{mrw1}. The contribution $\chi_j^2$ of the $j^{th}$ data set to the overall $\chi^2$ is given therein 
by Eq.~(1). The scale factors $z_j$, which minimise each $\chi_j^2$, are evaluated using Eq.~(2); the minimal $\chi_j^2$ value for each data set (denoted by $(\chi_j^2)_{min}$) is given in Eq.~(3) and the scaling contribution (of the $j^{th}$ data 
set) to $(\chi_j^2)_{min}$ in Eq.~(4). Finally, the scale factors for free floating $\hat{z}_j$ (which we will use in Section \ref{sec:Reproduction}, when investigating the absolute normalisation of the CX data using the ZUAS12 prediction as reference) 
are obtained from Eq.~(5); their total uncertainty $\Delta \hat{z}_j$ has been defined at the end of Section 2.2 of Ref.~\cite{mrw1}.

One statistical test will be performed on each data set, involving its contribution $(\chi_j^2)_{min}$ to the overall $\chi^2$. The corresponding p-value will be evaluated from the $(\chi_j^2)_{min}$ and the number of degrees of freedom of the data 
set (hereafter, the acronym DOF will stand for `degree(s) of freedom', whereas NDF will denote the `number of DOF'); for a data set with $N_j$ data points (none of which is an outlier), NDF is equal to $N_j$. Decisions on the tested data set will be 
made on the basis of the comparison of the corresponding p-value with the assumed confidence level $\mathrm{p}_{min}$. The value of $\mathrm{p}_{min}$ is fixed to the equivalent of a $2.5 \sigma$ effect in the normal distribution, corresponding to 
about $1.24 \cdot 10^{-2}$.

The repetitive referencing to the databases is largely facilitated if one adheres to the following notation: DB$_+$ for the $\pi^+ p$ database; DB$_-$ for the $\pi^- p$ elastic-scattering database; DB$_0$ for the CX database; DB$_{+/-}$ for the combined 
$\pi^\pm p$ elastic-scattering databases; DB$_{+/0}$ for the combined $\pi^+ p$ and CX databases. Furthermore, the prefix `t' (as, for instance, in tDB$_+$) denotes a `truncated' database, i.e., a database obtained after the removal of the outliers.

\section{\label{sec:CXData}The CX data}

The available measurements are listed (in the chronological order they had been reported) in Table \ref{tab:CSExp1}. During the last fifteen years, the database has been enlarged by a factor of seven, i.e., from a mere $47$ data points (which were 
available for the analyses \cite{glk,m}) to the present status of a total of $333$ data points. Of the added $286$ data points, $270$ relate to the differential cross section (DCS), $9$ to the total cross section (TCS), $6$ to the analysing power (AP), 
and $1$ data point to one threshold constant (to the isovector scattering length $b_1$).

A milestone in the low-energy CX experimentation was the FITZGERALD86 \cite{dhf} experiment, which took place at LAMPF and used the $\pi^0$ spectrometer to obtain (for the first time) important DCS data around the $s$- and $p$-wave interference minimum 
(see next section). By establishing a rigorous relation between the real parts of the $s$- and $p$-wave CX amplitudes at low energies, the FITZGERALD86 experiment became the backbone of the analyses \cite{glk,m} and was essential in terms of their 
conclusions on the violation of the isospin invariance in the hadronic part of the $\pi N$ interaction. The FITZGERALD86 experiment was the first complete CX experiment, as it also investigated (and reported) the normalisation uncertainty. Important in 
terms of the enhancement of the DB$_0$ was the ISENHOWER99 \cite{ldi} experiment, also performed at LAMPF with the $\pi^0$ spectrometer.

Within the last decade, the Crystal-Ball Collaboration made massive contributions to the low-energy DB$_0$ with experiments at the Brookhaven National Laboratory (BNL): the SADLER04 \cite{mesa} experiment added $60$ data points, whereas the 
successor of that experiment, the MEKTEROVI{\' C}09 \cite{mekt} experiment, contributed another $140$ data points. In total, the data obtained by the Crystal-Ball Collaboration amount now to $60 \%$ of the DB$_0$ which is available below $100$ MeV.

From the remaining experiments, the JIA08 \cite{jia} data have been taken around the $s$- and $p$-wave interference minimum, whereas the SCHROEDER01 \cite{ss} experimental result on the width of the $1s$ state of pionic hydrogen (corrected in 
Ref.~\cite{orwmg} after properly taking into account the contributions of the $\gamma n$ channel) led to the extraction of the scattering length $\tilde{a}^{c0}$ ($=\sqrt{2}b_1$). The FRLE{\v Z}98 \cite{efr} experiment investigated the angular distribution of 
the DCS at $27.50$ MeV. Finally, the BREITSCHOPF06 \cite{breit} experiment reported the TCSs at nine energies below $100$ MeV. The remaining experiments account for less than $10 \%$ of the DB$_0$ at low energies.

The complete DB$_0$ consists of $54$ data sets. The quoted values of the TCS in Refs.~\cite{smpr,bagh,mekt} have been extracted from the coefficients of the Legendre expansion of the angular distribution of the DCS and are thus correlated with the main 
results of these experiments; because of this correlation, one may use either set of values, but not both. We will use the Legendre coefficients of Refs.~\cite{smpr,bagh} (their DCS measurements have not been published), and directly the DCS data of 
Ref.~\cite{mekt}.

In our approach, all data sets must be accompanied by a normalisation uncertainty. This requirement also applies to one-point data sets, because the scale factors must be calculated in all cases (in order to enable the investigation of a possible bias 
in the analysis). As a result, realistic normalisation uncertainties had to be assigned to those experiments which did not report this quantity. We decided to assign these uncertainties as follows:
\begin{itemize}
\item $6 \%$ to BUGG71 \cite{bugg}, as this uncertainty was assigned to the experiments of the TCS or PTCS in Refs.~\cite{mrw1,mworg}.
\item $3 \%$ to BREITSCHOPF06, because the experimental group had already combined statistical and systematic effects in quadrature (and reported only the total uncertainty).
\item $3.1 \%$ to SALOMON84 \cite{smpr}, i.e., the normalisation uncertainty of the (similar, as well as close in time) BAGHERI88 \cite{bagh} experiment.
\item $8 \%$ to DUCLOS73 \cite{duc}, i.e., double the normalisation uncertainty of the ISENHOWER99 large-angle data sets.
\end{itemize}

The experimental results of Ref.~\cite{jia} contain asymmetric statistical uncertainties for the measured DCS. Unable to treat asymmetric uncertainties in the analysis, we will use the average of the (absolute values of the) two uncertainties for each 
input data point.

\section{\label{sec:IM}The $s$- and $p$-wave interference minimum}

The main contributions to the CX scattering amplitude in the low-energy region come from the real parts of the $s$ and $p$ waves. These contributions are of opposite signs and cancel each other in the forward direction around $45$ MeV. This 
destructive-interference phenomenon acts as a magnifying glass, probing the smaller contributions in the $\pi N$ dynamics, like those from the imaginary parts, from the $d$ and $f$ waves, and (potentially) from isospin breaking.

An estimate of the energy of the CX DCS minimum had been obtained in the FITZGERALD86 experiment, from a fit to the extrapolated DCS values to centre-of-mass (CM) scattering angle $\theta=0^\circ$. According to that estimate, the CX DCS minimum occurs 
at $T=45.0 \pm 0.5$ MeV; the minimal DCS value was also extracted: $\left( \frac{d\sigma}{d\Omega} \right) _{min}=2.4 \pm 0.5$ $\mu$b/sr \cite{dhf}. The ZUAS12 predictions~\footnote{The statistical uncertainty in the prediction for the energy $T$ of 
the CX DCS minimum was well below the quoted value. We opt for an increased uncertainty as we have not investigated the sensitivity of the extracted result to the variation of the $d$ and $f$ waves, which we have fixed herein (as in Ref.~\cite{mrw1}) 
from the current SAID solution (WI08) \cite{abws}.} are $43.7 \pm 0.1$ MeV and $0.3 \pm 0.1$ $\mu$b/sr, respectively.

\section{\label{sec:Fits_CX}Results of the fits to the CX measurements}

\subsection{\label{sec:K-Matrix_CX}Fits to the DB$_0$ using the $K$-matrix parameterisations}

For $T \leq 100$ MeV, each $K$-matrix element (i.e., each phase shift) can be approximated in terms of a finite number of energy-independent expansion parameters. The explicit forms of the parameterisations, used in the present work, have been given in 
Ref.~\cite{mrw1}. Since the presently-available experimental data in the energy domain of our analyses cannot determine the expansion coefficients of higher-order terms, we retain in our parametric forms only the terms up to $\mathcal{O}(\epsilon^2)$, 
where $\epsilon$ is the pion CM kinetic energy~\footnote{As in any phenomenological description of data, there is some arbitrariness in the choice of the forms used in our $K$-matrix parameterisations (e.g., using $\epsilon$ or the square of the CM 
momentum as expansion variable, expanding $K$ or $K^{-1}$, etc.). The chosen forms achieve the best reproduction of the experimental data, allowing simultaneously the determination of the fit parameters from the available $\pi N$ data below $100$ MeV. 
In general, the differences among the solutions, obtained with the forms which were examined, were found small (e.g., the differences in the resulting $\chi^2$ values were typically at the percent level).}. As when analysing the DB$_-$ in Section 3.2 
of Ref.~\cite{mrw1}, we will fix the $I=3/2$ amplitudes from the final fit to the tDB$_+$ using the $K$-matrix parameterisations (see Section 3.1 of that paper). The same $K$-matrix parameterisations for the $I=1/2$ amplitudes will be used in the 
description of the CX measurements, with different parameters $\tilde{a}_{0+}^{1/2}$, $b_1$, $c_1$, $d_{13}$, $e_{13}$, $d_{11}$, and $e_{11}$. (The fit parameter $b_1$ is not the standard isovector scattering length defined in Section \ref{sec:CXData}.)

The results of the optimisation procedure are shown in Table \ref{tab:OP}. Since seven parameters are used to generate the fitted values, the NDF in the first fit to the DB$_0$ was $326$; the minimum value of $\chi^2$ was $400.3$, indicating a rather 
coherent database. For the tDB$_0$, the minimum value of $\chi^2$ was $312.8$ for $321$ DOF in the fit. The optimal values of the parameters $\tilde{a}_{0+}^{1/2}$, $d_{13}$, and $e_{13}$ came out significantly different~\footnote{The differences 
between the two sets of values represent effects between $4.9$ and $7.4 \sigma$ in the normal distribution.} from those obtained in the analysis of the tDB$_-$ in Ref.~\cite{mrw1}. The details on the tDB$_0$, obtained from the final fit, are given in 
Table \ref{tab:DBCX}.

Although the results of Tables \ref{tab:OP} and \ref{tab:DBCX} provide ground for questioning the absolute normalisation of all seven FITZGERALD86 data sets, we decided to retain the absolute normalisation of the three remaining FITZGERALD86 data sets 
as their removal is not called for when strictly applying our rejection criteria. On the other hand, given the importance of FITZGERALD86 data in the analyses of Refs.~\cite{glk,m}, it goes without saying that the re-analysis of the CX measurements in 
terms of the violation of the isospin invariance in the hadronic part of the $\pi N$ interaction at low energies is imperative.

\subsection{\label{sec:Model}Common fit to the tDB$_{+/0}$ using the ETH model}

Details on the ETH model, as well as on its seven parameters ($G_\sigma$, $K_\sigma$, $G_\rho$, $K_\rho$, $g_{\pi NN}$, $g_{\pi N \Delta}$, and $Z$), may be obtained from Refs.~\cite{mrw1,mworg}. This isospin-invariant model was introduced in 
Ref.~\cite{glmbg} and was developed to its final form by the mid 1990s.

Prior to fitting the ETH model to the tDB$_{+/0}$, the data were subjected to a common fit using the $K$-matrix parameterisations. The tDB$_+$ consisted of $340$ data points (detailed in Table 1 of Ref.~\cite{mrw1}), whereas the tDB$_0$ comprised $328$ 
data points, i.e., the original $333$ data points minus the five outliers detailed in Table \ref{tab:OP} of this work. The common fit to these data using the $K$-matrix parameterisations resulted in the $\chi^2$ value of $737.0$ for $654$ DOF and no 
additional outliers.

The model fit to the data yielded the minimal $\chi^2$ value of $960.5$ for $661$ DOF. The optimal values of the model parameters from the fit to the tDB$_{+/0}$ are listed in Table \ref{tab:ModelParameters}; the uncertainties contain the Birge factor 
$\sqrt{\chi^2/\mathrm{NDF}}$, which takes account of the goodness of the fit. The table also contains the ZUAS12 solution for $\mathrm{p}_{min} \approx 1.24 \cdot 10^{-2}$. The differences between these two sets of values are evident, especially for 
$G_\rho$ and $g_{\pi NN}$. The correlation (Hessian) matrix, obtained in the fit, is given in Table \ref{tab:HessianMatrix}. 

We now reflect on the final $\chi^2$ values obtained so far. We first concentrate on the results for $\mathrm{p}_{min} \approx 1.24 \cdot 10^{-2}$. The separate fits to the data using the $K$-matrix parameterisations yielded the $\chi^2$ values of $427.2$, 
$371.0$ \cite{mrw1}, and $312.8$ (this work) for the tDB$_+$, tDB$_-$, and tDB$_0$, with $333$, $321$, and $321$ DOF, respectively. The $\chi^2$ values obtained with the $K$-matrix parameterisations in the two analyses of the combined truncated databases 
(i.e., tDB$_{+/-}$ and tDB$_{+/0}$) come out very close to the sum of the corresponding results for the separate fits: $792.4$ (instead of the sum of $798.1$) for the tDB$_{+/-}$ \cite{mrw1} and $737.0$ (instead of the sum of $740.0$) for the tDB$_{+/0}$ 
(this work). (As the NDF in the tDB$_-$ and tDB$_0$ are (by chance) identical, the results are directly comparable.) Therefore, we observe that, in the case of the fits using the $K$-matrix parameterisations, the difference of the two $\chi^2$ values 
(which is about $55.4$) reflects, almost entirely, the difference of the $\chi^2$ values in the separate fits to the tDB$_-$ and tDB$_0$ (which is equal to $58.1$, the smaller $\chi^2$ value for the fit to the tDB$_0$).

The increase of the $\chi^2$ values in the fits to the tDB$_{+/-}$ or to the tDB$_{+/0}$ using the ETH model (over the result of the fits to the same data using the $K$-matrix parameterisations) is due to the imposition of theoretical constraints (e.g., 
of the crossing and isospin symmetry); as earlier mentioned, the fits to the data using the $K$-matrix parameterisations are devoid of theoretical constraints, other than the expected low-energy behaviour of the $K$-matrix elements. All else being equal, 
one would expect that the difference in the $\chi^2$ values between the model fit to the data and the fit using the $K$-matrix parameterisations would (more or less) be the same for the two truncated databases, i.e., for the tDB$_{+/-}$ and tDB$_{+/0}$; 
however, this is far from being true. The value of $+55.4$ for the difference $\chi^2$(tDB$_{+/-}$)$-\chi^2$(tDB$_{+/0}$) in the fits using the $K$-matrix parameterisations turns into $-87.6$ with the use of the ETH model. This is the result of the 
considerably larger increase in the $\chi^2$ for the tDB$_{+/0}$ fits from the $K$-matrix parameterisations to the use of the ETH model: this increase amounts to $223.5$ compared to $80.5$ for the tDB$_{+/-}$ fits~\footnote{The importance of these 
differences may be easily assessed after considering that the variance of the $\chi^2$ distribution is equal to $2 \cdot \mathrm{NDF}$; therefore, the expectation for the `statistical fluctuation' in the quoted $\chi^2$ values is $\sqrt{2 \cdot \mathrm{NDF}}$.}. 
Evidently, the substitution of the tDB$_-$ with tDB$_0$ leads to a noticeable deterioration of the overall description of the experimental data in the model fits. This deterioration is a first indication of a general difficulty in the description of the 
tDB$_{+/0}$ in terms of \emph{one} set of parameter values of the ETH model. This fact can be explained if the theoretical basis upon which the data analysis rests (presumably, the isospin invariance in the hadronic part of the $\pi N$ interaction) is somewhat 
disturbed. Inspection of Table \ref{tab:ChiSquaresNDF} reveals that the information which is obtained from the results for the two other $\mathrm{p}_{min}$ levels used in Ref.~\cite{mrw1} (i.e., those corresponding to a $2$ and $3 \sigma$ effect in the 
normal distribution) matches very well the result for $\mathrm{p}_{min} \approx 1.24 \cdot 10^{-2}$. We will return to this issue in Section \ref{sec:Isospin}.

\subsubsection{\label{sec:ModelConstants}Threshold constants}

From the values of the model parameters and their uncertainties given in Table \ref{tab:ModelParameters}, as well as the correlation matrix given in Table \ref{tab:HessianMatrix}, we calculated the isoscalar and isovector $s$-wave scattering lengths and the 
isoscalar(isovector)-scalar(vector) $p$-wave scattering volumes. The results are:
\[
\frac{1}{3}\:\tilde{a}_{0+}^{1/2}+\frac{2}{3}\:\tilde{a}_{0+}^{3/2}=0.0059(25) \: \mu_c^{-1} ,
\]
\[
-\frac{1}{3}\:\tilde{a}_{0+}^{1/2}+\frac{1}{3}\:\tilde{a}_{0+}^{3/2}=-0.08245(56) \: \mu_c^{-1} , \]
\begin{equation}
\frac{1}{3}\:\tilde{a}_{1-}^{1/2}+\frac{2}{3}\:\tilde{a}_{1-}^{3/2}+\frac{2}{3}\:\tilde{a}_{1+}^{1/2}+\frac{4}{3}\:\tilde{a}_{1+}^{3/2}=0.2103(30) \: \mu_c^{-3} ,
\label{eq:atildas}
\end{equation}
\[
-\frac{1}{3}\:\tilde{a}_{1-}^{1/2}+\frac{1}{3}\:\tilde{a}_{1-}^{3/2}-\frac{2}{3}\:\tilde{a}_{1+}^{1/2}+\frac{2}{3}\:\tilde{a}_{1+}^{3/2}=0.1829(17) \: \mu_c^{-3} ,
\]
\[
\frac{1}{3}\:\tilde{a}_{1-}^{1/2}+\frac{2}{3}\:\tilde{a}_{1-}^{3/2}-\frac{1}{3}\:\tilde{a}_{1+}^{1/2}-\frac{2}{3}\:\tilde{a}_{1+}^{3/2}=-0.1940(18) \: \mu_c^{-3} ,
\]
\[
-\frac{1}{3}\:\tilde{a}_{1-}^{1/2}+\frac{1}{3}\:\tilde{a}_{1-}^{3/2}+\frac{1}{3}\:\tilde{a}_{1+}^{1/2}-\frac{1}{3}\:\tilde{a}_{1+}^{3/2}=-0.0697(11) \: \mu_c^{-3} .
\]

Converting these results to the standard spin-isospin quantities, we obtain
\[
\tilde{a}_{0+}^{3/2}=-0.0765(25) \: \mu_c^{-1} ,\qquad \tilde{a}_{0+}^{1/2}=0.1708(29) \: \mu_c^{-1} ,
\]
\begin{equation}
\tilde{a}_{1+}^{3/2}=0.2190(19) \: \mu_c^{-3} ,\qquad \tilde{a}_{1+}^{1/2}=-0.0337(12) \: \mu_c^{-3} ,
\label{eq:atildasvalues}
\end{equation}
\[
\tilde{a}_{1-}^{3/2}=-0.0447(13) \: \mu_c^{-3} ,\qquad \tilde{a}_{1-}^{1/2}=-0.0883(22) \: \mu_c^{-3} .
\]
Significant differences are found when comparing the values of $\tilde{a}_{0+}^{1/2}$, $\tilde{a}_{1+}^{3/2}$, and $\tilde{a}_{1-}^{1/2}$ with the corresponding results of Ref.~\cite{mrw1}.

From the results in Eqs.~(\ref{eq:atildasvalues}), we obtain
\[
\tilde{a}^{cc} = \frac{2}{3} \, \tilde{a}_{0+}^{1/2} + \frac{1}{3} \, \tilde{a}_{0+}^{3/2} = 0.0884(26) \: \mu_c^{-1} \,
\]
and
\[
\tilde{a}^{c0} = \sqrt{2} \left( -\frac{1}{3}\:\tilde{a}_{0+}^{1/2}+\frac{1}{3}\:\tilde{a}_{0+}^{3/2} \right) = -0.11660(79) \: \mu_c^{-1} \, .
\]
Unlike the $\tilde{a}^{cc}$ value extracted from the tDB$_{+/-}$ \cite{mrw1}, the value of the present work is compatible with the result of the measurement of the strong shift of the $1s$ state of pionic hydrogen \cite{ss}. Additionally, 
the value of $\tilde{a}^{c0}$ is marginally consistent (the difference between the two values is at the level of $1.9 \sigma$) with the result of the same experiment for the width of the $1s$ state of pionic hydrogen. (In this comparison, 
the em corrections of Ref.~\cite{orwmg} have been applied to the raw experimental results of Ref.~\cite{ss}.)

\subsubsection{\label{sec:ModelPhases}Hadronic phase shifts}

\begin{figure}
\begin{center}
\includegraphics [width=15.5cm] {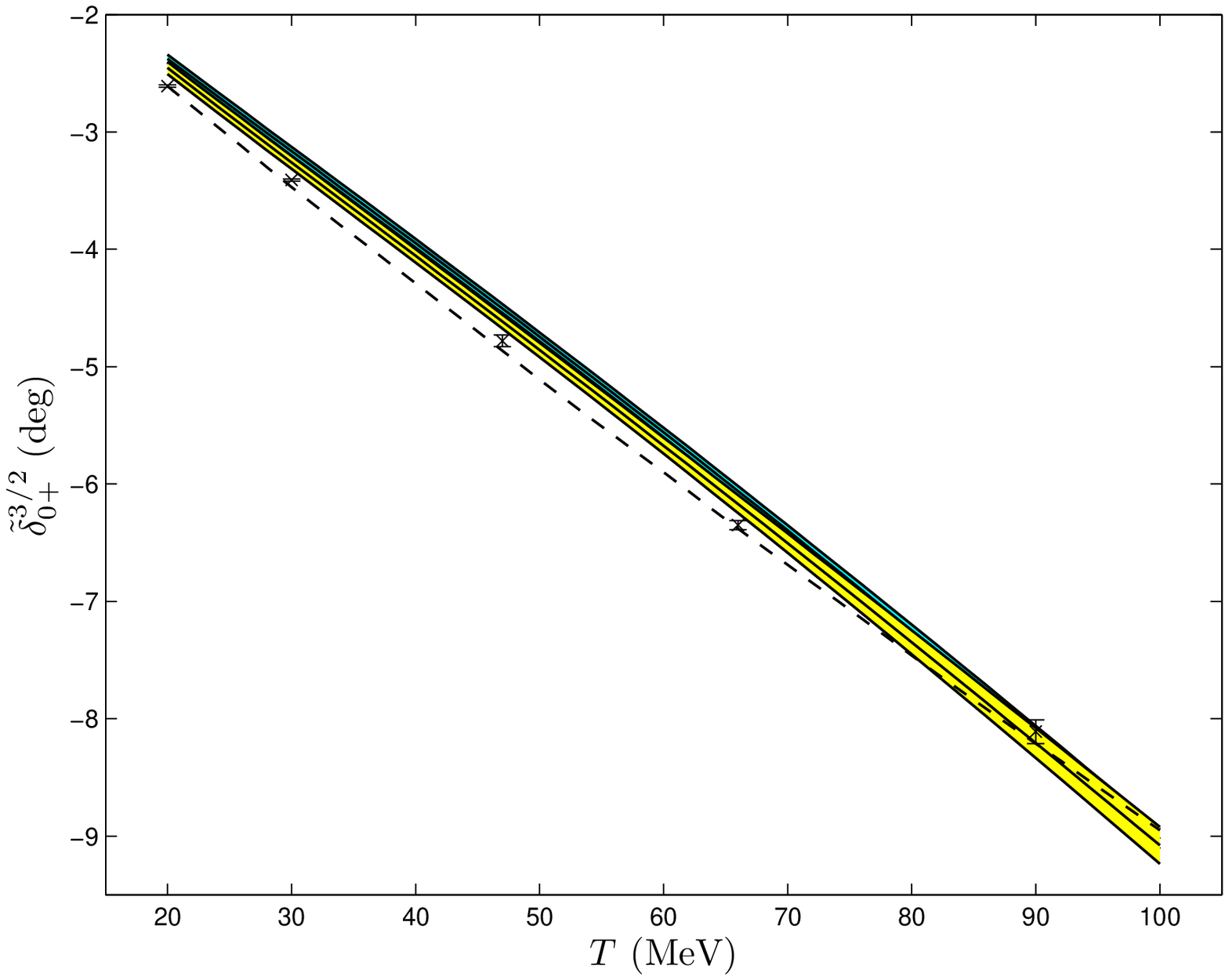}
\caption{\label{fig:a}The energy dependence of the em-modified hadronic phase shift $\tilde{\delta}_{0+}^{3/2}$ (S31) from the present work, along with $1 \sigma$ uncertainties (yellow band). Also included is the ZUAS12 prediction 
\cite{mrw1}, obtained on the basis of the elastic-scattering data below $100$ MeV, along with the corresponding $1 \sigma$ uncertainties (blue band). The current SAID solution (WI08) \cite{abws} is represented by the dashed curve; 
the five points shown (at $T=20$, $30$, $47$, $66$, and $90$ MeV) are the single-energy WI08 values.}
\end{center}
\end{figure}

\begin{figure}
\begin{center}
\includegraphics [width=15.5cm] {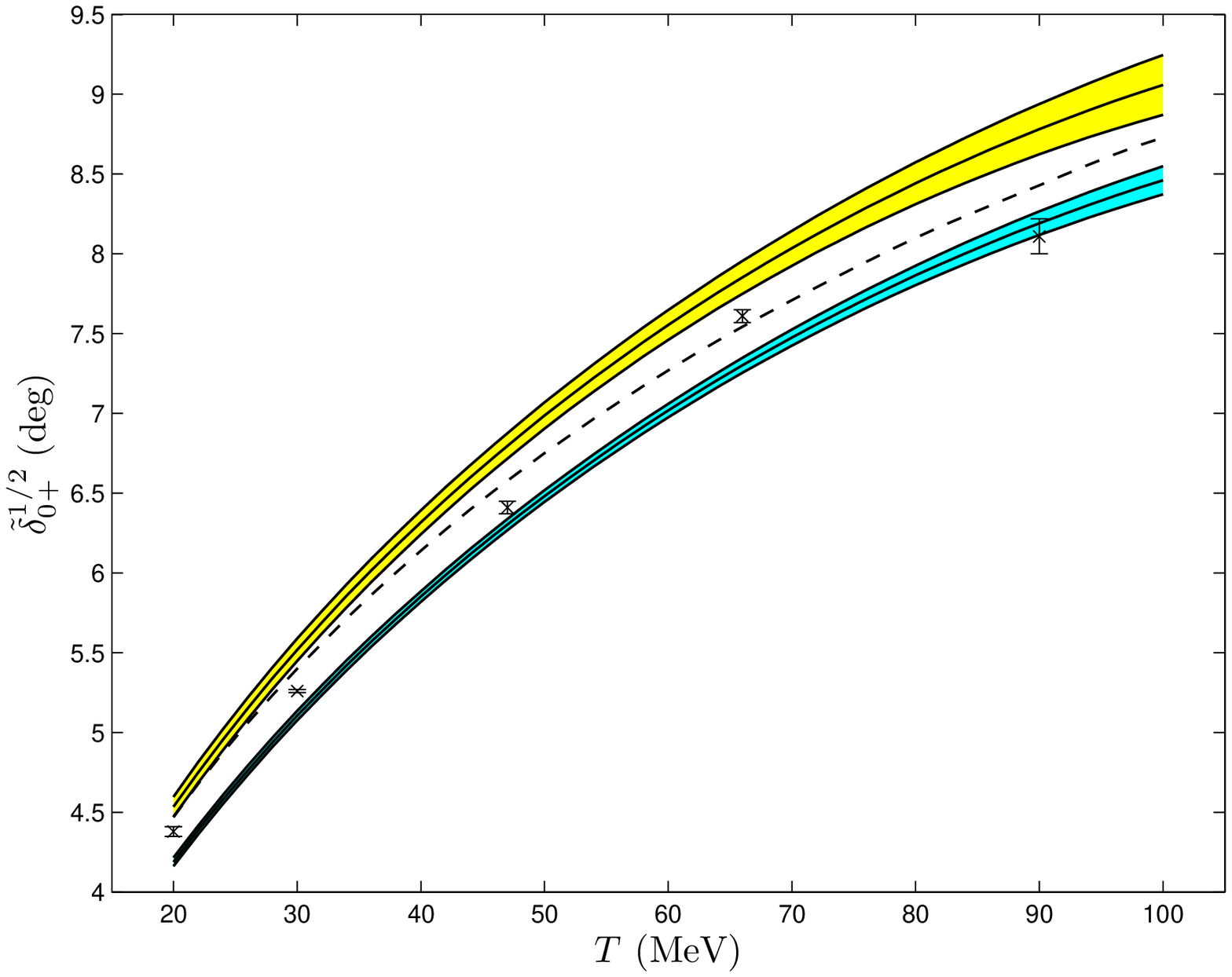}
\caption{\label{fig:b}The energy dependence of the em-modified hadronic phase shift $\tilde{\delta}_{0+}^{1/2}$ (S11) from the present work, along with $1 \sigma$ uncertainties (yellow band). Also included is the ZUAS12 prediction 
\cite{mrw1}, obtained on the basis of the elastic-scattering data below $100$ MeV, along with the corresponding $1 \sigma$ uncertainties (blue band). The current SAID solution (WI08) \cite{abws} is represented by the dashed curve; 
the five points shown (at $T=20$, $30$, $47$, $66$, and $90$ MeV) are the single-energy WI08 values.}
\end{center}
\end{figure}

\begin{figure}
\begin{center}
\includegraphics [width=15.5cm] {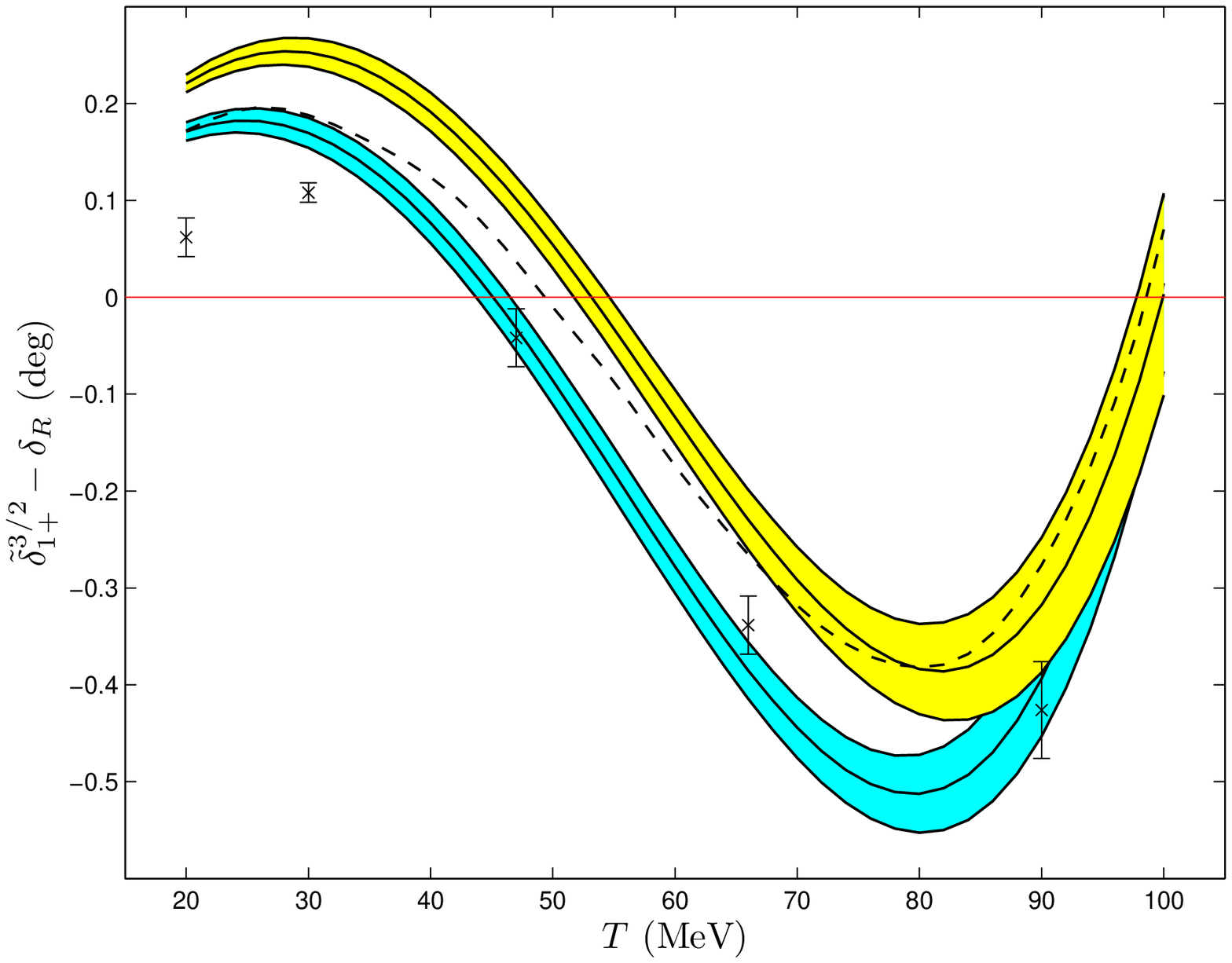}
\caption{\label{fig:c}The energy dependence of the em-modified hadronic phase shift $\tilde{\delta}_{1+}^{3/2}$ (P33) from the present work, along with $1 \sigma$ uncertainties (yellow band). Also included is the ZUAS12 prediction 
\cite{mrw1}, obtained on the basis of the elastic-scattering data below $100$ MeV, along with the corresponding $1 \sigma$ uncertainties (blue band). The current SAID solution (WI08) \cite{abws} is represented by the dashed curve; 
the five points shown (at $T=20$, $30$, $47$, $66$, and $90$ MeV) are the single-energy WI08 values. To enable a better comparison of the values contained in this figure, an energy-dependent baseline $\delta_R$ 
($=(0.20 \cdot T+1.54)T \cdot 10^{-2}$, with $T$ in MeV and $\delta_R$ in degrees) has been subtracted from all data.}
\end{center}
\end{figure}

\begin{figure}
\begin{center}
\includegraphics [width=15.5cm] {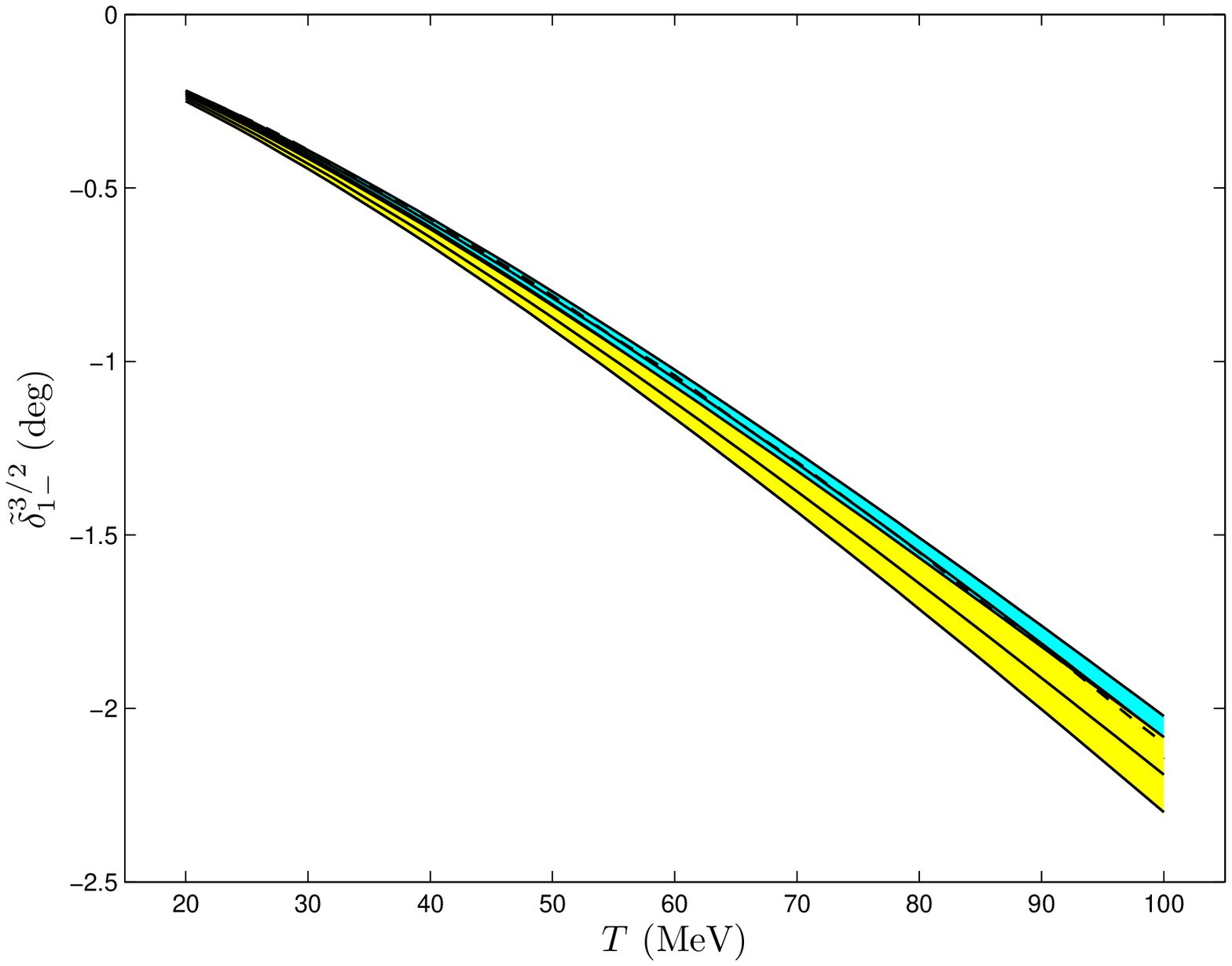}
\caption{\label{fig:d}The energy dependence of the em-modified hadronic phase shift $\tilde{\delta}_{1-}^{3/2}$ (P31) from the present work, along with $1 \sigma$ uncertainties (yellow band). Also included is the ZUAS12 prediction 
\cite{mrw1}, obtained on the basis of the elastic-scattering data below $100$ MeV, along with the corresponding $1 \sigma$ uncertainties (blue band). The current SAID solution (WI08) \cite{abws} is represented by the dashed curve.}
\end{center}
\end{figure}

\begin{figure}
\begin{center}
\includegraphics [width=15.5cm] {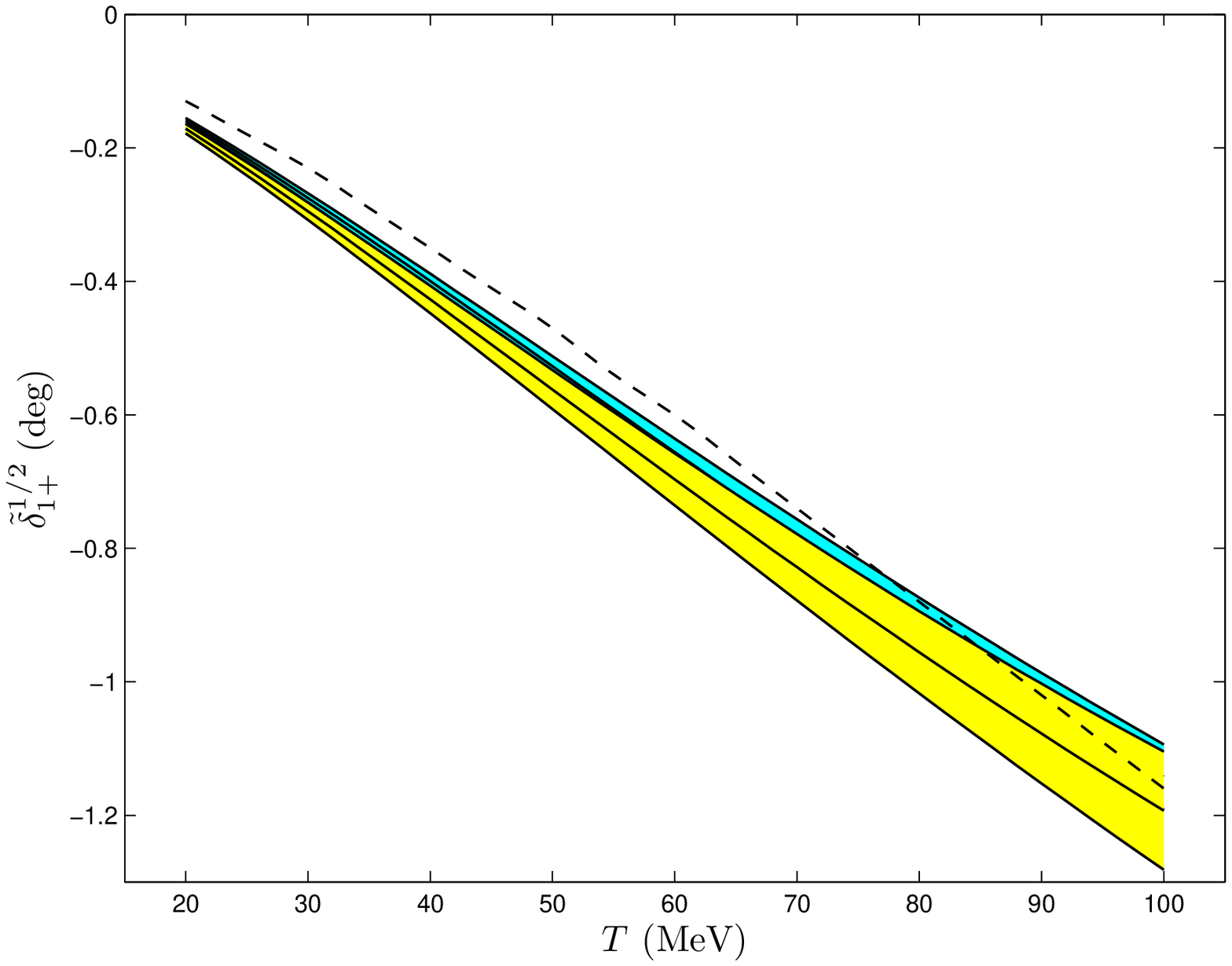}
\caption{\label{fig:e}The energy dependence of the em-modified hadronic phase shift $\tilde{\delta}_{1+}^{1/2}$ (P13) from the present work, along with $1 \sigma$ uncertainties (yellow band). Also included is the ZUAS12 prediction 
\cite{mrw1}, obtained on the basis of the elastic-scattering data below $100$ MeV, along with the corresponding $1 \sigma$ uncertainties (blue band). The current SAID solution (WI08) \cite{abws} is represented by the dashed curve.}
\end{center}
\end{figure}

\begin{figure}
\begin{center}
\includegraphics [width=15.5cm] {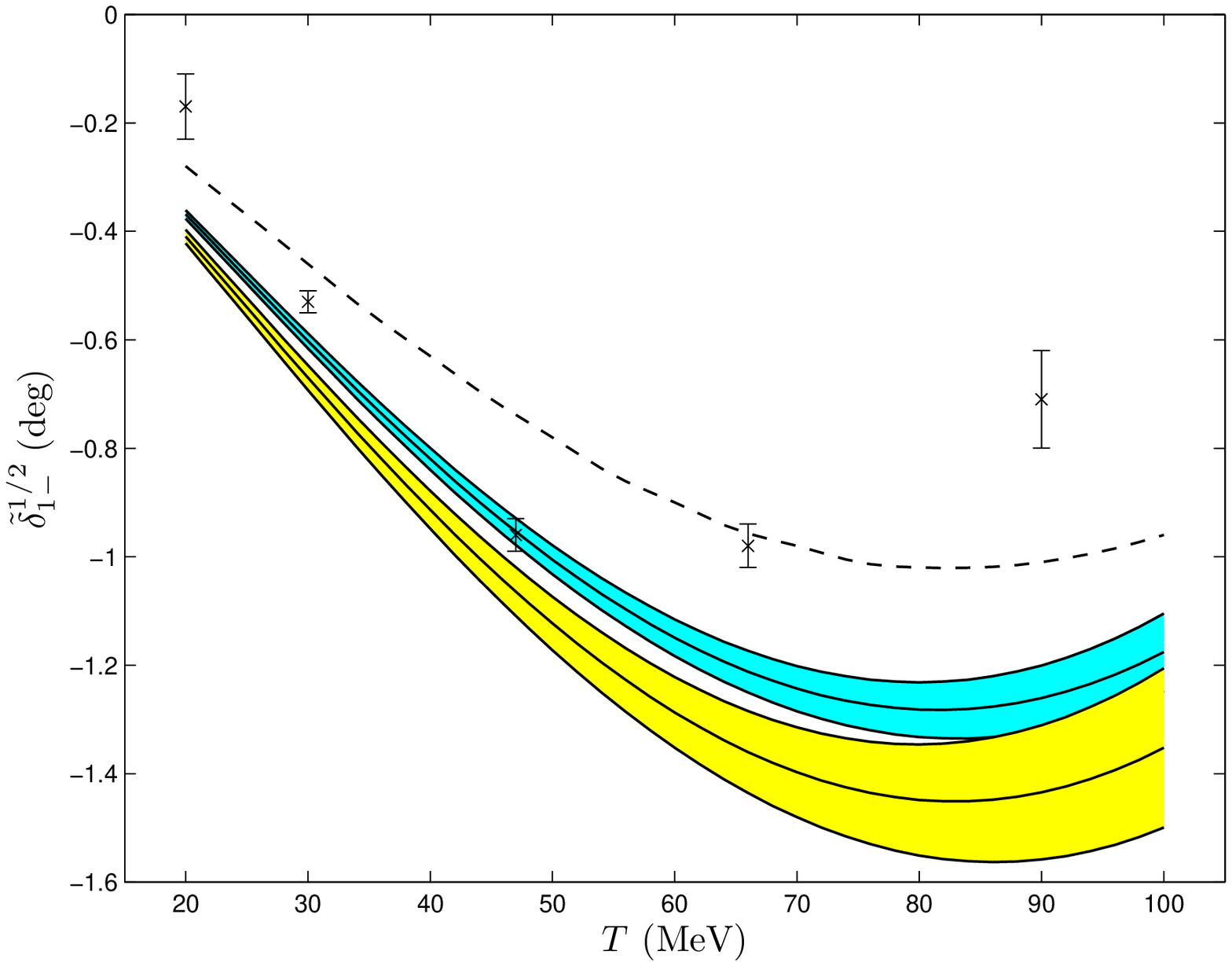}
\caption{\label{fig:f}The energy dependence of the em-modified hadronic phase shift $\tilde{\delta}_{1-}^{1/2}$ (P11) from the present work, along with $1 \sigma$ uncertainties (yellow band). Also included is the ZUAS12 prediction 
\cite{mrw1}, obtained on the basis of the elastic-scattering data below $100$ MeV, along with the corresponding $1 \sigma$ uncertainties (blue band). The current SAID solution (WI08) \cite{abws} is represented by the dashed curve; 
the five points shown (at $T=20$, $30$, $47$, $66$, and $90$ MeV) are the single-energy WI08 values.}
\end{center}
\end{figure}

The results for the $s$- and $p$-wave phase shifts, from the fit to the tDB$_{+/0}$ using the ETH model, are given in Table \ref{tab:PhaseShifts}. These hadronic phase shifts are also shown in Figs.~\ref{fig:a}-\ref{fig:f}, together with the 
ZUAS12 results, as well as the current SAID solution (WI08) \cite{abws} and their five single-energy values (wherever available). A very noticeable difference is seen in the case of $\tilde{\delta}_{0+}^{1/2}$. Smaller differences may be seen 
in two $p$-wave phase shifts, i.e., in $\tilde{\delta}_{1+}^{3/2}$ and $\tilde{\delta}_{1-}^{1/2}$.

\subsubsection{\label{sec:ScaleFactors}Scale factors and normalised residuals}

Similarly to the tests performed in Section 3.4.4 of Ref.~\cite{mrw1}, we will first investigate whether any bias is present in the distribution of the scale factors $z_j$, extracted in the final step of the optimisation scheme. Subsequently, we 
will address the issue of the distribution of the normalised residuals of the fit.

\begin{figure}
\begin{center}
\includegraphics [width=15.5cm] {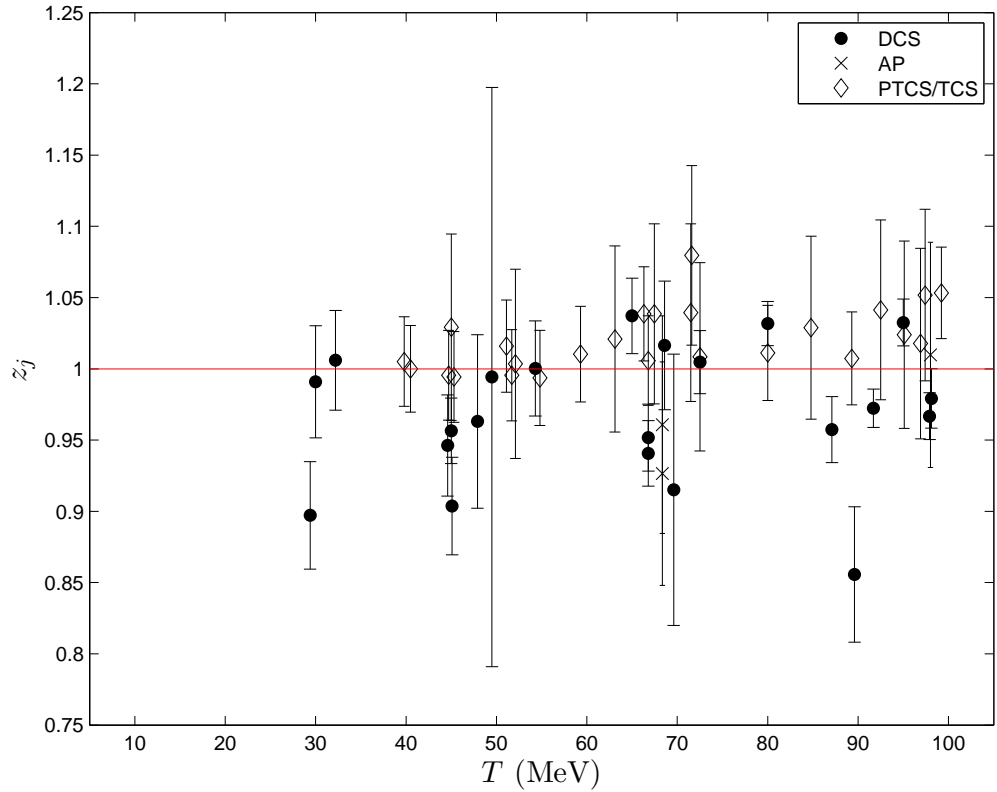}
\caption{\label{fig:sfpip}The scale factors $z_j$ of the $\pi^+ p$ data sets, obtained from the common fit to the truncated $\pi^+ p$ and CX databases using the ETH model (see Section \ref{sec:Model}). The values, corresponding 
to the two data sets which were freely floated (see Table 1 of Ref.~\cite{mrw1}), have not been included. The results of the linear fit to the shown values are given in Section \ref{sec:ScaleFactors}.}
\end{center}
\end{figure}

\begin{figure}
\begin{center}
\includegraphics [width=15.5cm] {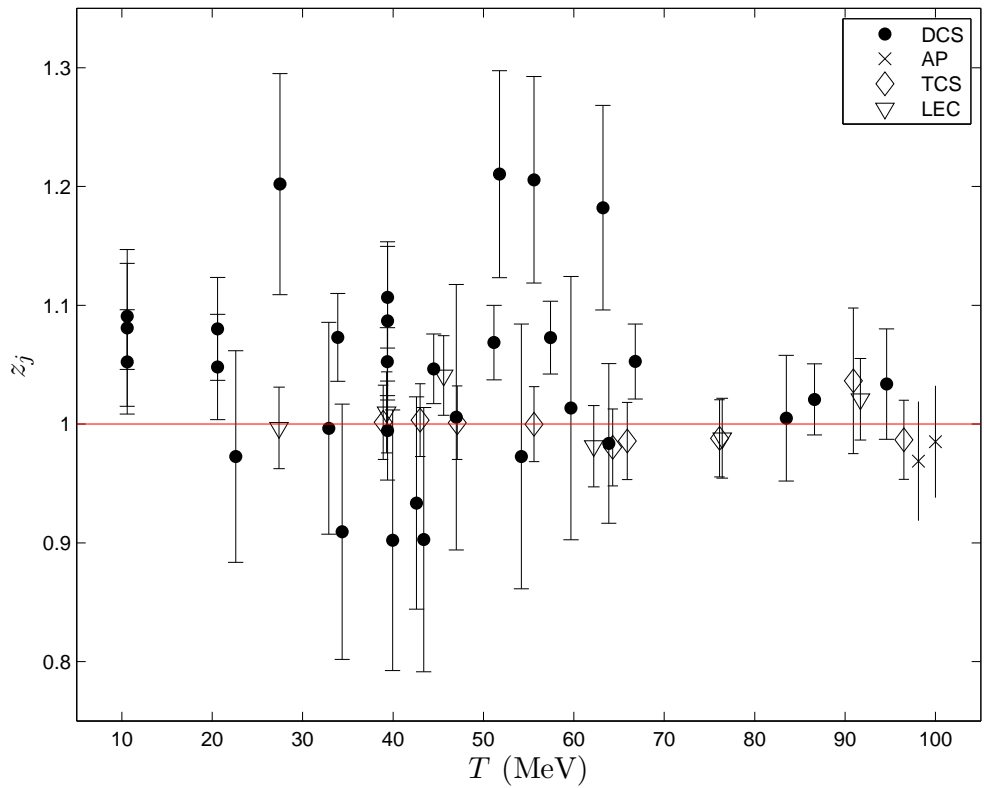}
\caption{\label{fig:sfpim}The scale factors $z_j$ of the CX data sets, obtained from the common fit to the truncated $\pi^+ p$ and CX databases using the ETH model (see Section \ref{sec:Model}). The values, corresponding to the 
four data sets which were freely floated (see Table \ref{tab:DBCX}), have not been included. The results of linear fits to the shown values are given in Section \ref{sec:ScaleFactors}. The data sets with the largest scale factors 
$z_j$ are the three remaining FITZGERALD86 data sets, as well as the FRLE{\v Z}98 data set.}
\end{center}
\end{figure}

Two linear fits (the pion laboratory kinetic energy being the independent variable) to the optimal $z_j$ values of Figs.~\ref{fig:sfpip} (for the $\pi^+ p$) and \ref{fig:sfpim} (for the CX reaction) were performed. The results of these two fits 
do not match well. The two intercept values were: $0.972 \pm 0.019$ in the case of the $\pi^+ p$ reaction and $1.062 \pm 0.016$ for the CX reaction. The slope was found to be compatible with $0$ in the former case: $(2.6 \pm 2.5) \cdot 10^{-4}$ 
MeV$^{-1}$. A noticeable energy dependence was found in the latter; the slope value came out equal to $(-6.8 \pm 2.7) \cdot 10^{-4}$ MeV$^{-1}$.

In Section \ref{sec:K-Matrix_CX}, we questioned the determination of the absolute normalisation in the FITZGERALD86 experiment, yet allowed three of these data sets to take part in the optimisation (as our criteria for rejection were not 
fulfilled). The fitted values of the intercept and slope are almost left intact in case the scale factors of these three data sets are not included in the linear fit examining the energy dependence of the scale factors; the intercept value 
came out equal to $1.060 \pm 0.014$, whereas the slope $(-6.9 \pm 2.4) \cdot 10^{-4}$ MeV$^{-1}$. It may also be argued that the extracted values of the intercept and of the slope show sensitivity to the inclusion in the fit of the three 
low-energy ($10.60$ MeV) entries of the ISENHOWER99 experiment (which we have no reason to question). We removed these three entries from the input and repeated the fit (also continuing to exclude the three afore-mentioned FITZGERALD86 
entries); the intercept value came out equal to $1.053 \pm 0.016$, whereas the slope $(-5.9 \pm 2.7) \cdot 10^{-4}$ MeV$^{-1}$.

These results establish a rather problematic situation (from the point of view of the analysis of the measurements with the ETH model) when forcing the data of these two reactions into a common optimisation scheme; the different values of 
the two intercepts demonstrate the overall tendency in the optimisation, with the generation of \emph{overestimated} fitted DCS values for the $\pi^+ p$ reaction and \emph{underestimated} ones for the CX reaction. It appears that the 
optimisation of the description of the input data is achieved at the expense of creating a bias in the reproduction of the two subsets (reactions) comprising the set of the input measurements. Equivalently, one might claim that the $I=3/2$ 
amplitudes obtained with the model have a difficulty to simultaneously account for the $\pi^+ p$ and CX reactions. As these difficulties were not present in the PSA of the $\pi^\pm p$ elastic-scattering data (at least, at a noticeable level), 
one may pose the question whether it makes sense to include the CX measurements into a common optimisation scheme, along with data from other reactions.

\begin{figure}
\begin{center}
\includegraphics [width=15.5cm] {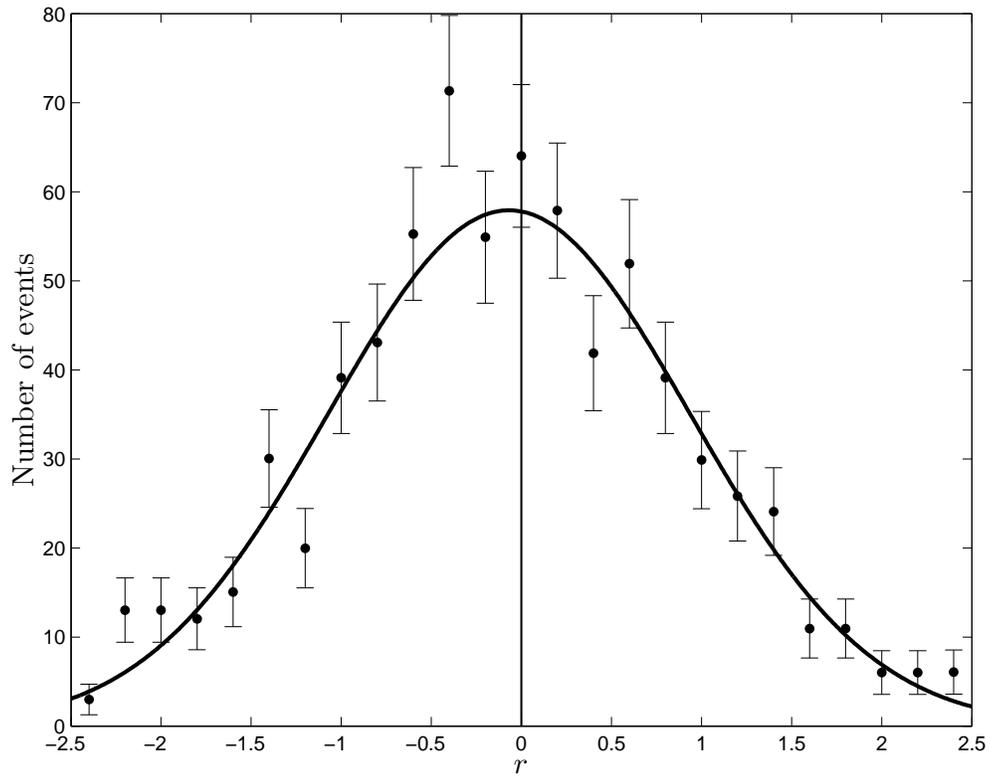}
\caption{\label{fig:residuals}The distribution of the normalised residuals, obtained from the common fit to the truncated $\pi^+ p$ and CX databases using the ETH model (see Section \ref{sec:ScaleFactors}). Also shown is the optimal 
Gaussian fit to the data (solid curve).}
\end{center}
\end{figure}

The distribution of the normalised residuals is shown in Fig.~\ref{fig:residuals}, along with the optimal Gaussian function; the $\chi^2$ value of this fit was $25.5$ for $22$ DOF. The offset $\bar{r}$ (for the definition, see Section 3.4.4 
of Ref.~\cite{mrw1}) came out equal to $(-6.8 \pm 4.2) \cdot 10^{-2}$. For the sake of completeness, we also give the optimal value and the uncertainty of parameter $B$ of the Gaussian fit to the data: $B=0.497 \pm 0.038$; the expectation 
value for $B$ is $0.5$.

\subsubsection{\label{sec:TCS}Reproduction of the $\pi^- p$ PTCSs}

We mentioned in Section 3 of Ref.~\cite{mrw1} that, as the nine existing $\pi^- p$ PTCSs and total-nuclear cross sections contain a component from CX scattering, they could not have been used in our PSA of the elastic-scattering data; we 
added that the inclusion of these data in any part of that analysis would perplex the discussion on the violation of the isospin invariance in the hadronic part of the $\pi N$ interaction. The results of the reproduction of the 
three~\footnote{Concerning the FRIEDMAN90 measurement \cite{frn}, we are aware of the revision in the energy calibration of the M11 pion channel at TRIUMF (which took place in the early 1990s), yet we have not found another published value 
for this measurement; the corrected energy values for the $\pi^+ p$ PTCSs of Ref.~\cite{frn} appeared a few years after the original publication.} measured PTCSs (where the contributions to the observable may be easily disentangled) are 
shown in Table \ref{tab:ReproductionOfPTCS}. We notice that the contributions from the CX reaction are large and substantiate our decision to avoid including these data in the fits to the elastic-scattering measurements.

The measurements of the $\pi^- p$ PTCS may be compared to the results obtained after summing up the contribution of the $\pi^- p$ elastic-scattering PTCS and that of the entire CX TCS; this is dictated by the experimental technique employed 
in these measurements, namely, the detection of only the $\pi^-$'s (interacting or passing through) downstream of the target, within a cone of aperture $2 \theta_L$ with its apex at the geometrical centre of the target, where $\theta_L$ is 
the laboratory-angle cut associated with the measurement ($30^\circ$ for the three available data points). Regarding the component of the CX TCS, one may use either the prediction from the PSA of the two elastic-scattering reactions \cite{mrw1} 
or the one obtained from the results of the present work; the better reproduction of the experimental data when invoking the CX TCS of the present work would be in favour of this paper. Although a slight preference for the results of this work 
has been seen, the experimental uncertainties are too large to lead to definite conclusions.

\section{\label{sec:Reproduction}Investigation of the absolute normalisation of the CX data on the basis of the ZUAS12 prediction}

We will next investigate the absolute normalisation of the CX data sets using the ZUAS12 prediction as reference. To this end, we must determine the amount at which the reference predictions for each CX data set (i.e., the $y_{ij}^{th}$ values 
appearing in Eq.~(1) of Ref.~\cite{mrw1}) must be floated in order to optimally reproduce the experimental data of the specific CX data set (i.e., the $y_{ij}^{exp}$ values). Therefore, relevant in this part of the study are the scale factors 
for free floating $\hat{z}_j$, given in Eq.~(5) of Ref.~\cite{mrw1}.

\begin{figure}
\begin{center}
\includegraphics [width=15.5cm] {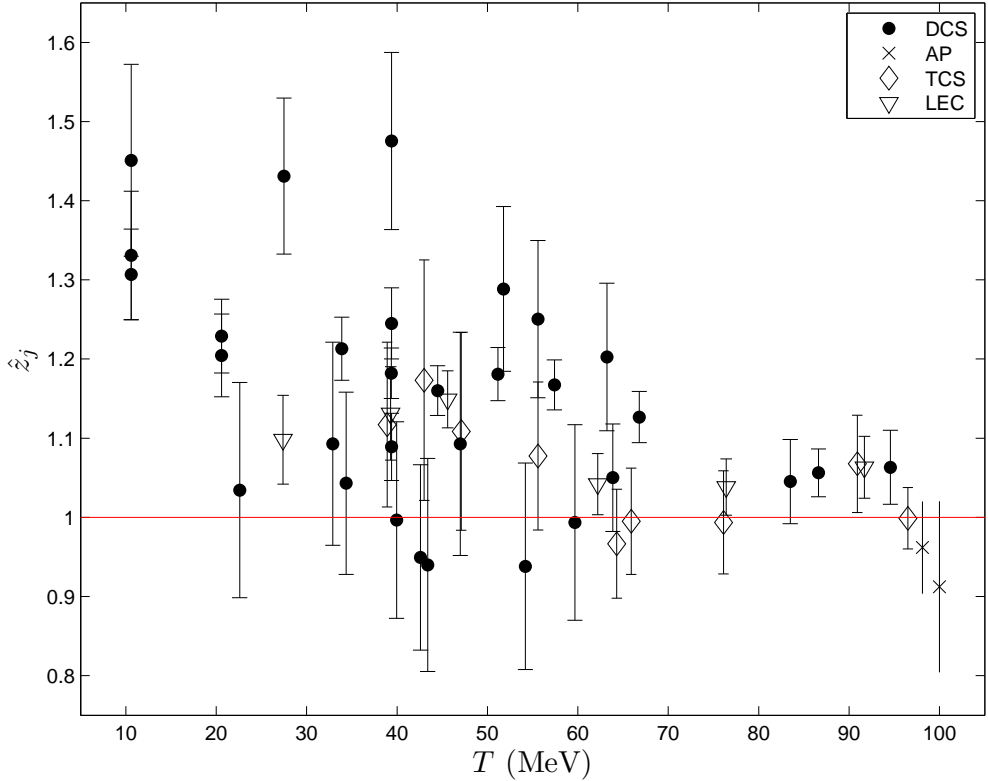}
\caption{\label{fig:sfpimCX}The scale factors $\hat{z}_j$ for free floating (evaluated with Eq.(5) of Ref.~\cite{mrw1}) of the CX data sets, obtained on the basis of the ZUAS12 predictions \cite{mrw1}, plotted separately for differential 
cross sections (DCS), total cross sections (TCS), analysing powers (AP), and the results for the coefficients of the Legendre expansion of the DCS (LEC). The four FITZGERALD86 data sets, which had been freely floated (see Table \ref{tab:DBCX}), 
as well as the BREITSCHOPF06 $75.10$ MeV entry, have not been included. Not shown in the figure is also the result of Ref.~\cite{ss} for the isovector scattering length $b_1$.}
\end{center}
\end{figure}

The extracted values of the scale factors $\hat{z}_j$ and their total uncertainties may be found in Table \ref{tab:CSExp2} and, plotted separately for the DCS, TCS, AP, and LEC measurements, in Fig.~\ref{fig:sfpimCX}. The four FITZGERALD86 data 
sets which had been freely floated in Section \ref{sec:K-Matrix_CX} are not shown; their $\hat{z}_j$ factors came out equal to $2.03(14)$, $2.26(15)$, $2.28(14)$, and $1.52(13)$ (the order corresponds to increasing energy). Even when using only 
the CX data, the scale factors obtained for these data sets (see Table \ref{tab:DBCX}) significantly exceed the expectation value of $1$. The $\hat{z}_j$ factor of the BREITSCHOPF06 one-point data set which was eliminated in Section 
\ref{sec:K-Matrix_CX} came out equal to $0.927(50)$; this data point is also not shown in Fig.~\ref{fig:sfpimCX}. One additional data point, the scattering length $b_1$ of Ref.~\cite{ss}, has not been included in this figure. In principle, one 
could assign this data point to the DCS set, in which case the outcome would have been consistent with the scale factors for free floating of the ISENHOWER99 $10.60$ MeV data; nevertheless, only genuine DCS measurements are shown in Fig.~\ref{fig:sfpimCX}.

Inspection of Fig.~\ref{fig:sfpimCX} leaves no doubt that, when using the ZUAS12 prediction as reference, the CX scale factors for free floating contain a large amount of fluctuation. As the ZUAS12 prediction is smooth, the fluctuation observed 
in the figure reflects the scattering of the absolute normalisation of the CX data sets. For instance, the $\hat{z}_j$ value for the FRLE{\v Z}98 data set comes out equal to $1.431(99)$. This data set lies in between three data sets with 
considerably smaller $\hat{z}_j$ values, i.e., between the two ISENHOWER99 $20.60$ MeV data sets and the MEKTEROVI{\' C}09 $33.89$ MeV data set. The values of the absolute normalisation of the two neighbouring data sets of DUCLOS73 ($22.60$ and 
$32.90$ MeV), as well as that of the JIA08 $34.37$ MeV data set, are compatible with the reference prediction.

Restricting ourselves below $70$ MeV, we calculated the weighted average of the $\hat{z}_j$ factors of Table \ref{tab:CSExp2} and Fig.~\ref{fig:sfpimCX}. The result is that the low-energy CX measurements lie on average $(15.6 \pm 1.4) \%$ above 
the ZUAS12 predictions~\footnote{The exclusion of the three remaining FITZGERALD86 data sets leaves this result almost intact.}. Naively translated into a relative difference in the CX scattering amplitude, this result would be equivalent to an effect 
around the $7-8 \%$ level.

In order to provide some perspective and motivation to research groups which are active in the low-energy $\pi N$ experimentation, we will now give the predictions obtained on the basis of the ZUAS12 and ZUAS12a solutions in a number of situations. 
We will investigate the differences in these two prediction sets and identify the kinematical regions which provide fertile ground for distinguishing experimentally between the two sets of values. We commence with the DCSs and TCSs. (The results 
on the Legendre-expansion coefficients are expected to follow the sensitivity of the DCS.)

\begin{figure}
\begin{center}
\includegraphics [width=15.5cm] {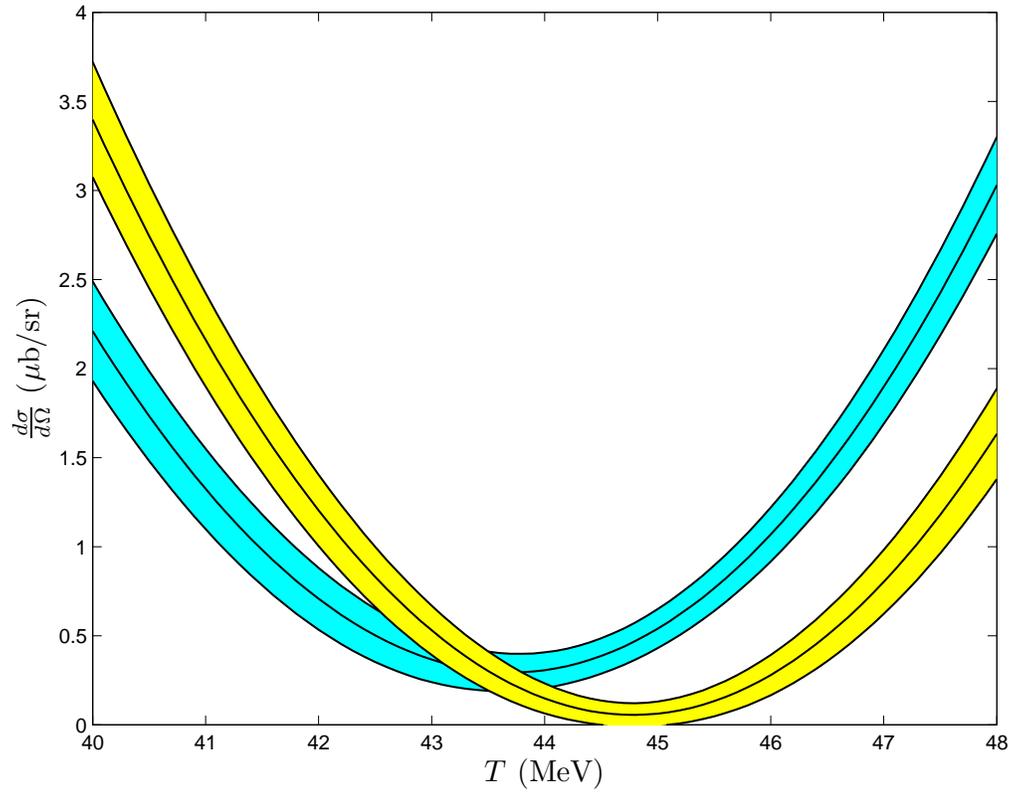}
\caption{\label{fig:IM}Two predictions for the CX DCS for CM scattering angle $\theta=0^\circ$ around the $s$- and $p$-wave interference minimum. The ZUAS12 prediction \cite{mrw1} is represented by the blue band, whereas 
the ZUAS12a one (this work) by the yellow band. Both bands indicate $1\sigma$ uncertainties.}
\end{center}
\end{figure}

The two predictions around the CX DCS minimum are shown in Fig.~\ref{fig:IM} for $\theta=0^\circ$. We observe that the two predictions differ; the ZUAS12a prediction exceeds the one obtained on the basis of the ZUAS12 results by about $1.1$ MeV. 
Additionally, the ZUAS12a solution predicts a deeper DCS minimum.

\begin{figure}
\begin{center}
\includegraphics [width=15.5cm] {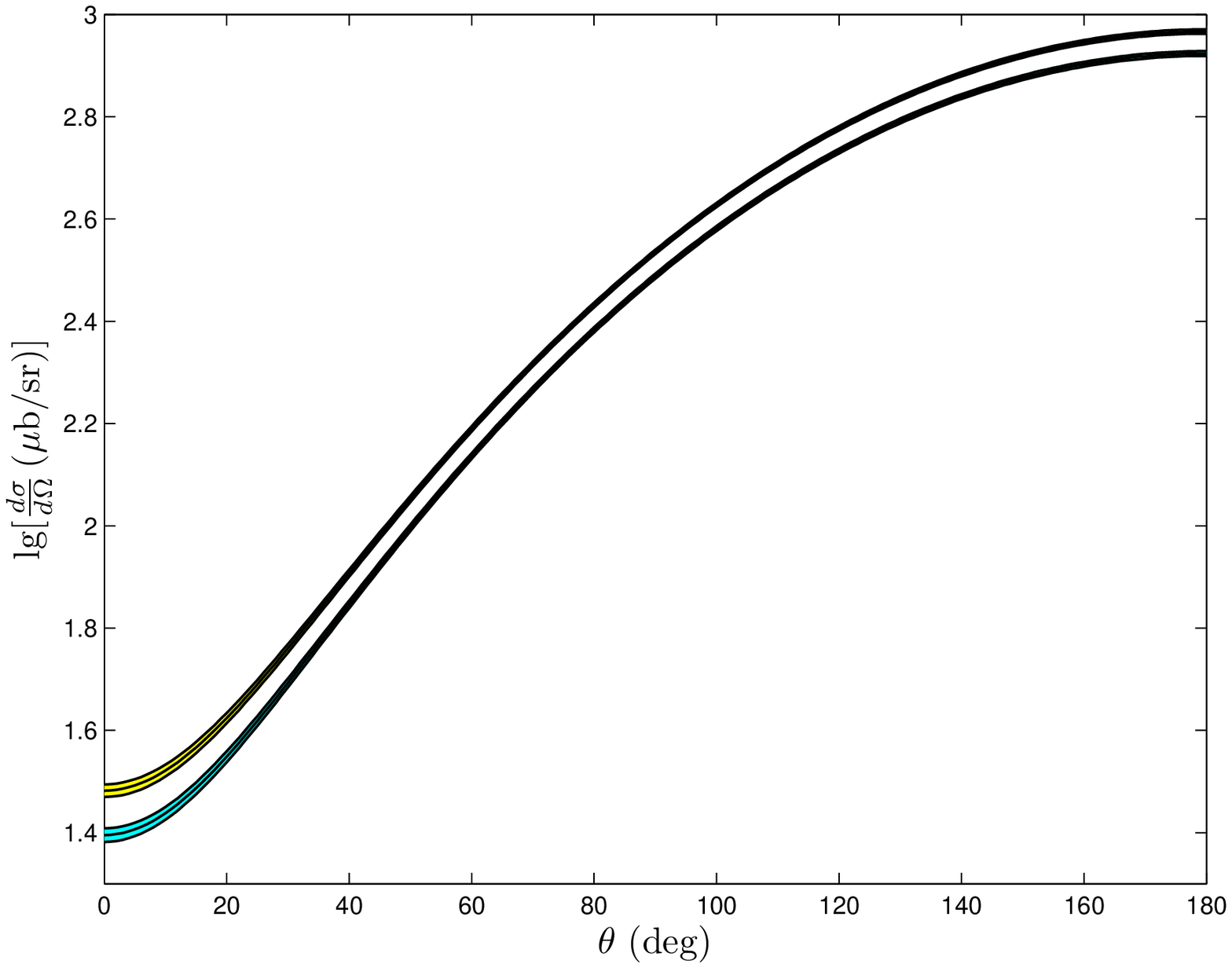}
\caption{\label{fig:CXDCS30MeV}Two predictions for the angular distribution of the CX DCS at $30$ MeV; $\theta$ denotes the CM scattering angle. The ZUAS12 prediction \cite{mrw1} is represented by the blue band, whereas the 
ZUAS12a one (this work) by the yellow band. Both bands indicate $1\sigma$ uncertainties.}
\end{center}
\end{figure}

The shapes of the angular distributions of the CX DCS, obtained on the basis of the ZUAS12 and ZUAS12a solutions, are different below and above the $s$- and $p$-wave interference minimum. Below the minimum, the ZUAS12a-based DCS systematically 
exceeds the ZUAS12 predictions, by varying amounts; at $20$ MeV, the relative difference (i.e., the difference normalised to the corresponding ZUAS12 values) is $16.5 \%$ for $\theta=0^\circ$ decreasing to $11.3 \%$ at $\theta=180^\circ$; the 
corresponding numbers for $30$ MeV are: $22.1$ and $10.4 \%$; finally, at $40$ MeV, the relative differences are: $53.8$ and $9.4 \%$. Evidently, the relative difference between the two predictions in the forward direction increases as the beam 
energy approaches the energy of the $s$- and $p$-wave interference minimum. Large effects are also seen in the TCS, slightly decreasing with increasing energy, from about $12.2 \%$ at $20$ MeV, to $11.2 \%$ at $30$ MeV, and to $10.0 \%$ at $40$ 
MeV. A representative plot of the two predictions for the angular distribution of the CX DCS below the $s$- and $p$-wave interference minimum is shown in Fig.~\ref{fig:CXDCS30MeV} for $30$ MeV.

\begin{figure}
\begin{center}
\includegraphics [width=15.5cm] {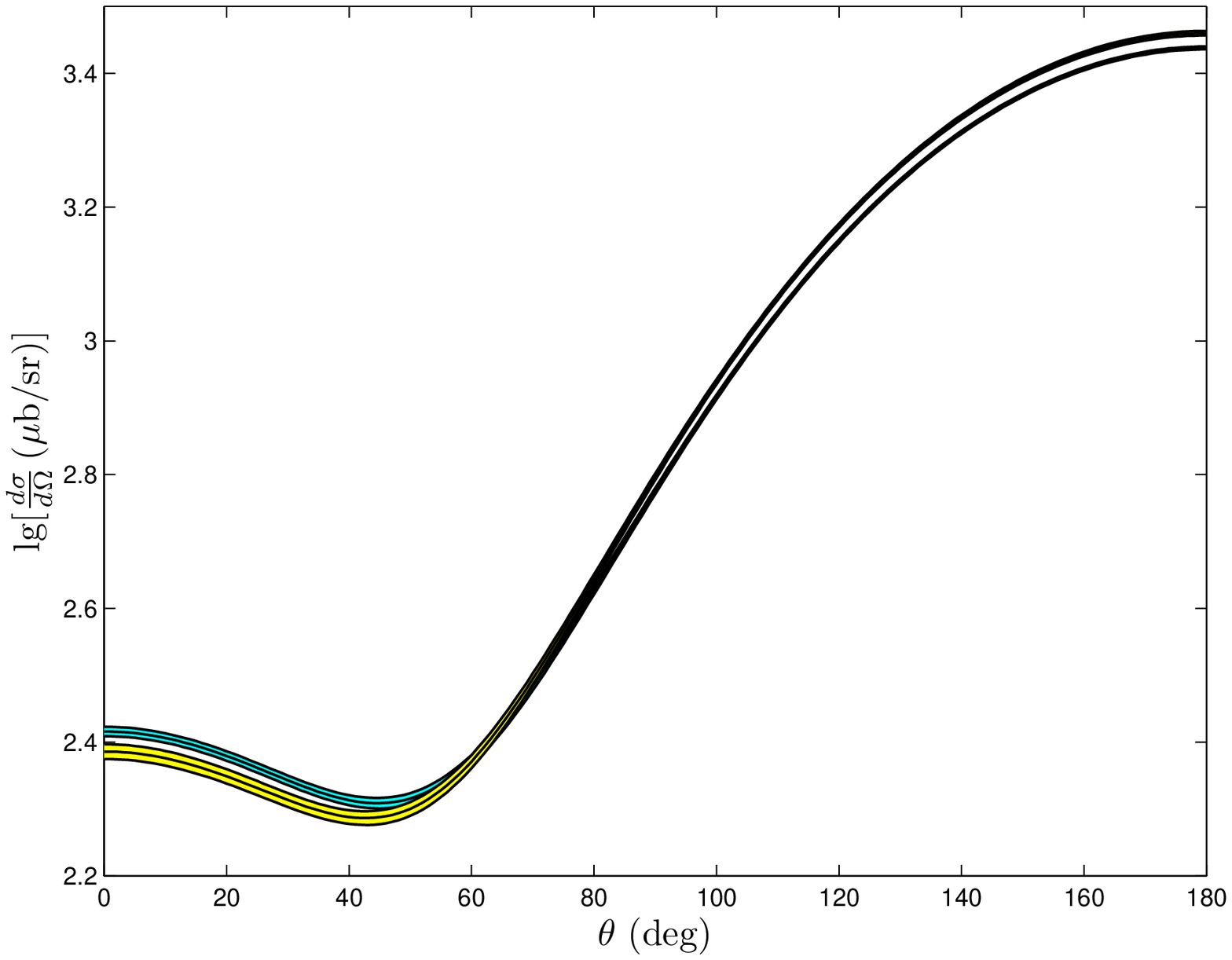}
\caption{\label{fig:CXDCS80MeV}Two predictions for the angular distribution of the CX DCS at $80$ MeV; $\theta$ denotes the CM scattering angle. The ZUAS12 prediction \cite{mrw1} is represented by the blue band, whereas the 
ZUAS12a one (this work) by the yellow band. Both bands indicate $1\sigma$ uncertainties.}
\end{center}
\end{figure}

The picture is slightly different at the high end of the energies: the ZUAS12a result lies below the ZUAS12 prediction in the forward direction. The two predictions cross each other between $40$ and $80^\circ$; above the crossing, the ZUAS12a 
prediction exceeds the ZUAS12 one (by smaller amounts when compared to the low energies). At $60$ MeV, the relative difference between the two predictions at $\theta=0^\circ$ is $-11.7 \%$, increasing to $7.3 \%$ at $\theta=180^\circ$; at $80$ 
MeV, the values are: $-6.7$ and $5.2 \%$, whereas at $100$ MeV, they are: $-6.8$ and $3.4 \%$. The net effect in the TCS decreases with increasing energy, from about $7.3 \%$ at $60$ MeV, to $4.6 \%$ at $80$ MeV, and to $2.1 \%$ at $100$ MeV. A 
representative plot of the two predictions for the angular distribution of the CX DCS above the $s$- and $p$-wave interference minimum is shown in Fig.~\ref{fig:CXDCS80MeV} for $80$ MeV.

The AP shows high sensitivity to the effect under investigation around the $s$- and $p$-wave interference minimum. The few available measurements of the AP in the low-energy CX reaction have been taken at $98.10$ and $100.00$ MeV, 
where the differences between the two prediction sets are small.

\begin{figure}
\begin{center}
\includegraphics [width=15.5cm] {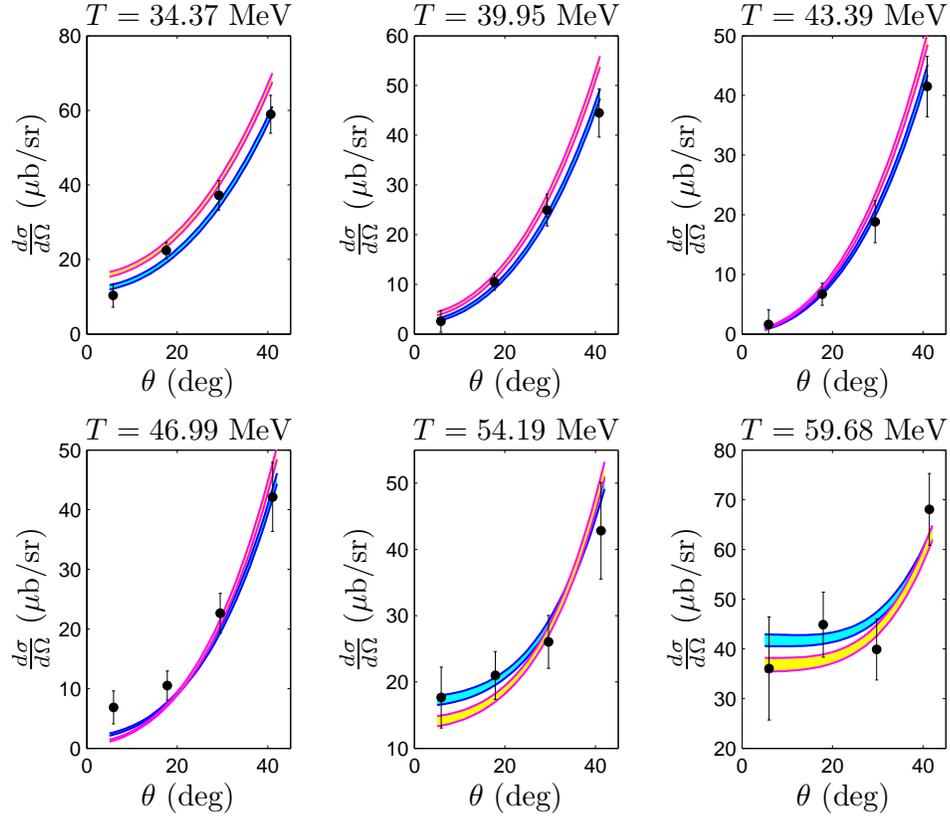}
\caption{\label{fig:JIA08}Two predictions for the angular distribution of the CX DCS in the kinematical region of the JIA08 \cite{jia} experiment. The ZUAS12 prediction \cite{mrw1} is represented by the blue band, whereas the 
ZUAS12a one (this work) by the yellow band. Both bands indicate $1\sigma$ uncertainties. Only the statistical uncertainties of the measurements of Ref.~\cite{jia} are shown (i.e., the $10 \%$ normalisation uncertainty of the 
experiment has not been included).}
\end{center}
\end{figure}

We finally comment on the experiment of Ref.~\cite{jia}, which took data at forward angles, at six energies around the CX DCS minimum. The authors have made the point that their data `show no evidence for unexpected isospin-breaking effects'. To 
start with, according to Table \ref{tab:DBCX} and assuming the correctness of the absolute normalisation of the bulk of the CX data, the JIA08 measurements lie between $0.5$ and $1.5 \sigma$ ($\sigma$, in this case, being equivalent to $10 \%$) 
below the optimal solution obtained \emph{only} from the CX data in Section \ref{sec:K-Matrix_CX}. To investigate the issue further, we generated the ZUAS12 and ZUAS12a predictions, corresponding to the values of the energy and CM scattering angle 
of these data. As seen in Figs.~\ref{fig:JIA08}, these two predictions lie close to one another in the kinematical region of the measurements. As a result, the experiment indeed agrees with the ZUAS12 prediction, but it also does with the ZUAS12a 
prediction (which was not available at the time the report of the experimental group appeared). Given the large normalisation uncertainties of the low-energy $\pi N$ experiments, as well as the general closeness of the ZUAS12 and ZUAS12a predictions, 
it is rather unlikely that any single experiment can disprove the violation of the isospin invariance in the hadronic part of the $\pi N$ interaction, especially one with a normalisation uncertainty at the $10 \%$ level.

\section{\label{sec:Causes}Possible causes of the observed differences between the ZUAS12 and ZUAS12a predictions}

We now summarise the main results obtained so far.
\begin{itemize}
\item Three of the values of the parameters of the $K$-matrix parameterisations (i.e., $\tilde{a}_{0+}^{1/2}$, $d_{13}$, and $e_{13}$), extracted in the analysis of the tDB$_{+/0}$, differ significantly from those obtained in the analysis of the 
tDB$_{+/-}$.
\item Two of the values of the parameters of the ETH model (i.e., $G_\rho$ and $g_{\pi NN}$), extracted in the analysis of the tDB$_{+/0}$, differ significantly from those obtained in the analysis of the tDB$_{+/-}$.
\item When using the ETH model, the substitution of the tDB$_-$ with the tDB$_0$ leads to noticeable deterioration of the results of the fits, indicating difficulties in the description of these measurements on the basis of \emph{one} set of 
parameter values of the ETH model. In this respect, the results of Table \ref{tab:ChiSquaresNDF}, for the three values of $\mathrm{p}_{min}$ which have been used in Ref.~\cite{mrw1}, are consistent.
\item A significant difference has been seen in the $s$-wave phase shift $\tilde{\delta}_{0+}^{1/2}$; smaller differences are observed in two $p$-wave phase shifts, i.e., in $\tilde{\delta}_{1+}^{3/2}$ and in $\tilde{\delta}_{1-}^{1/2}$.
\item The reproduction of the absolute normalisation of the CX data sets on the basis of the ZUAS12 prediction is poor below $70$ MeV. The differences observed amount to a $7-8 \%$ effect in the CX scattering amplitude.
\end{itemize}
These differences between the ZUAS12 and ZUAS12a solutions and the predictions obtained on their basis are significant. We will now attempt to identify possible causes of the observed discrepancies.

In view of the results obtained so far, there are three assumptions, of which at most two can be simultaneously valid.
\begin{itemize}
\item The absolute normalisation of the bulk of the low-energy $\pi N$ experimental data is reliable.
\item The residual contributions in the em corrections, applied to the experimental data in order to extract the hadronic part of the $\pi N$ amplitude, are negligible.
\item The isospin invariance in the hadronic part of the $\pi N$ interaction holds.
\end{itemize}
We will next elaborate on each of the three possibilities arising from the nonfulfillment of the aforementioned presumptions.

\subsection{\label{sec:IntegrityDB}Experimental problems}

The first explanation for the observed differences involves a trivial effect, namely the systematic incorrectness of the absolute normalisation in the low-energy $\pi N$ data. Arguments have been presented in Ref.~\cite{mworg}, to substantiate 
the claim that some of the reported uncertainties in the $\pi N$ experimentation have been underestimated; concerning this last issue, the data analysis dictates that the two elastic-scattering reactions are more affected than the CX.

Our first point concerns the statistical uncertainties of the data points. When visually inspecting the low-energy $\pi N$ data, one is frequently unable to comprehend how it is possible for successive measurements (i.e., at neighbouring values 
of the CM scattering angle) to be so different. There are many occasions in the databases where the statistical uncertainties of the data points seem questionable.

Our second point concerns the systematic effects. It is not understood how the absolute normalisation of some experiments, e.g., of the FITZGERALD86 data sets at the three lowest energies, may be wrong (according to the bulk of the tDB$_0$) by 
an average of about $70 \%$ (and of a fourth data set by $45 \%$), at a time when the reported normalisation uncertainty in the experiment was $7.8 \%$. Such an effect may only be caused by any of three reasons (or their combination): a) the 
determination of the absolute normalisation in the experiment had been erroneous, b) the energy of the incoming beam had not been what the experimenters expected, or c) the normalisation uncertainty in the experiment had been grossly underestimated.

Our third point concerns the small values of the normalisation uncertainty reported in many low-energy $\pi N$ experiments; for instance, the reported normalisation uncertainties in $37$ out of the $90$ data sets in the initial DB$_{+/-}$ are 
below $3 \%$. One could possibly try to modify the small values of the normalisation uncertainty (perhaps, by setting a lower limit at $3 \%$), yet such an approach seems arbitrary. Consequently, one is left with no other option than to rely on 
an approach which places importance on the absolute normalisation of the bulk of the experimental data and to apply a reasonable procedure for the elimination of the outliers; this is the approach which we set up in Refs.~\cite{mrw1,mrw2,mworg}.

There is one disturbing discrepancy in the entire analysis of the tDB$_{+/-}$ which cannot be easily put aside, namely, the disagreement between the $\tilde{a}^{cc}$ value obtained as an extrapolation from the experimental data (above the $\pi N$ 
threshold) and the one extracted directly at the $\pi N$ threshold from the strong shift of the $1s$ level in pionic hydrogen. Assuming the correctness of both the absolute normalisation of the bulk of the elastic-scattering databases and of the 
raw measurement of $\epsilon_{1s}$ \cite{ss}, the two extracted values should be compatible, if a consistent set of em corrections (i.e., corrections which have been obtained within the same framework and which are also complete in the sense of 
containing the contributions from all relevant physical effects) have been applied to the raw data. In case of important residual effects in the em corrections (see next section), the question surfaces as to the energy dependence of these 
contributions.

\subsection{\label{sec:EMCorrections}Residual contributions in the em corrections}

Although it is not clear how these contributions can modify so drastically the overall picture and especially the results of Table \ref{tab:ChiSquaresNDF} (any residual em contributions are expected to affect equally the description of the experimental 
data on the basis of the $K$-matrix parameterisations and of the ETH model), the completeness of the em corrections in the $\pi N$ system at low energies is an important issue which must be properly defined and treated. In Refs.~\cite{mworg,orwmg}, 
some details are given on the effects which the stage-II em corrections contain; these effects are mostly related to the use of the physical instead of the (unknown) hadronic masses for the proton, the neutron, and the charged and neutral pion in the 
determination of the em corrections \cite{gmorw1,gmorw2}. On the other hand, it may be argued that the appropriate inclusion of these effects should lead to the optimal description of the input data. If this is the case, then the iterative procedure which 
had been set forth in the determination of the em corrections in Refs.~\cite{gmorw1,gmorw2} must have captured some of these effects. Unable, however, to either validate this statement or refute it, we can only encourage the theoretical re-assessment of the 
role of the em effects in the low-energy $\pi N$ interaction.

\subsection{\label{sec:Isospin}The violation of the isospin invariance in the hadronic part of the $\pi N$ interaction}

This is the last of the options which may be put forth in an attempt to explain the observed discrepancies and, admittedly, the most interesting one in Physics terms. This possibility may account for the results of Table \ref{tab:ChiSquaresNDF}. 
There has been a great amount of discussion regarding the acceptance of the conclusions of Refs.~\cite{glk,m}. One is tempted, however, to pose the question: `Why should the isospin invariance in the $\pi N$ system be obeyed in the first place?' 
After all, the hadronic masses of the $u$ and $d$ quarks are different; similarly, the masses of the nucleons differ (beyond trivial em effects), and so do those of the $\Delta$'s. It appears, therefore, that the appropriate question to ask is not 
whether the isospin invariance is violated, but at which amount it is. Within the framework of the heavy-baryon Chiral-Perturbation Theory, the group of Mei\ss{}ner have repeatedly treated isospin-breaking effects in the $\pi N$ system at low 
energies and concluded that the discrepancy should be at most at the percent level (e.g., see Ref.~\cite{meis} and the references cited therein).

Two mechanisms had been proposed in the past, to account for the violation of the isospin invariance in the hadronic part of the $\pi N$ interaction at the level of Feynman graphs: the first mechanism affects the elastic scattering ($\rho^0 - \omega$ 
mixing \cite{coon}-\cite{m1}), the second the CX reaction ($\eta - \pi^0$ mixing \cite{cutk}). As both the $\omega$ and the $\eta$ states are singlets, the coupling of the former to the $\rho^0$ and of the latter to the $\pi^0$ explicitly violate the isospin 
invariance. Given that, in the case of the elastic scattering, only one graph (i.e., the $t$-channel $\rho$-exchange graph) is affected, whereas in the case of the CX reaction all graphs are influenced (see Fig.~\ref{fig:IsospinBreakingEtaPi0}), one 
would be prone to conclude that the isospin-breaking effects are more important in the latter case; however, Ref.~\cite{cutk} concluded with the statement that `\dots the isospin violation from $\eta - \pi^0$ mixing can be safely ignored in $\pi N$ 
partial-wave analyses.' In fact, the possibility of large isospin-breaking effects in specific kinematical regions is not refuted in Ref.~\cite{cutk}; for instance, emphasis in that paper was placed on the effects induced in the amplitude 
of one higher baryon resonance, which were expected to be around the $7 \%$ level. Unfortunately, there is no indication that the kinematical region around the $\pi N$ threshold received equal attention in Ref.~\cite{cutk}. Of course, this is not 
very surprising given the scarcity of the low-energy $\pi N$ measurements around the time Ref.~\cite{cutk} appeared; in fact, below $T=100$ MeV, the only CX DCS measurements, which were available at that time, were the $3$ data points of Ref.~\cite{duc}.

We noticed that the coupling constant $g_{\pi N N}$ is significantly affected when substituting the tDB$_-$ with the tDB$_0$. Of course, if the isospin invariance is violated, there is not \emph{one} coupling constant $g_{\pi N N}$; one must distinguish 
between $g_{\pi^0 p p}$, $g_{\pi^0 n n}$, and $g_{\pi^\pm p n}$. In this case, the fits to the elastic-scattering data essentially determine $g_{\pi^\pm p n}$, whereas those involving the CX reaction also contain contributions from $g_{\pi^0 p p}$ and 
$g_{\pi^0 n n}$. As a result, the value of the coupling constant $g_{\pi N N}$ extracted from the common fits to the tDB$_{+/0}$ represents a weighted average of these three quantities. The differences observed imply that at least one of the two 
$g_{\pi^0 N N}$ coupling constants differs from $g_{\pi^\pm p n}$. The isospin-breaking effects on $g_{\pi N N}$ have been studied theoretically and generally found to be small; for instance, Ref.~\cite{mh} evaluated these effects using QCD 
sum rules and reported that $g_{\pi^\pm p n}$ should be equal to the average of the two $g_{\pi^0 N N}$ values and that the splitting should be expected between $1.2$ and $3.7 \%$. The difference between the two $g_{\pi N N}$ values of Table 
\ref{tab:ModelParameters} is larger, around the $4.5 \%$ level.

\section{\label{sec:Discussion}Discussion and Summary}

This study concludes the analysis of the presently-available pion-nucleon ($\pi N$) data below pion laboratory kinetic energy of $100$ MeV. The separate analysis of the data for the two elastic-scattering and for the charge-exchange (CX) reactions 
was enabled via suitable parameterisations of the $s$- and $p$-wave $K$-matrix elements at low energies. Common fits to the data were performed using either these $K$-matrix parameterisations or the ETH model of Ref.~\cite{glmbg}, which is based on 
meson-exchange $t$-channel graphs, as well as on the $s$- and $u$-channel $N$ and $\Delta$ contributions. The analysis with the $K$-matrix parameterisations led to the identification of the outliers in the databases and tested the self-consistency 
of the input prior to its submission to the fits using the ETH model. The optimal values of the model parameters, as well as their correlation matrix, were obtained from the ensuing fits and were used as the basis for generating Monte-Carlo predictions 
for the em-modified hadronic phase shifts, for the low-energy $\pi N$ constants, as well as for the standard observables (i.e., for the differential cross section (DCS), analysing power (AP), partial-total cross section (PTCS), and total cross section 
(TCS)) for any of the three reactions, at any value of the relevant kinematical variables (i.e., energy and scattering angle for the DCS and AP, energy and laboratory-angle cut for the PTCS, and energy for the TCS).

Given that the electromagnetic (em) corrections (which are applied to the hadronic phase shifts in order to extract the $\pi N$ partial-wave amplitudes, which, in turn, lead to the observables) of Refs.~\cite{gmorw1,gmorw2} have been obtained by using 
the physical, instead of the (unknown) hadronic, masses for the proton, the neutron, and the charged and neutral pion, we have assumed the cautious attitude of considering the various physical quantities (i.e., the model parameters, the low-energy $\pi N$ 
constants, the hadronic phase shifts, etc.) not purely hadronic, but em-modified hadronic. At the present time, one cannot assess the importance of the residual em effects (i.e., of the stage-II em corrections).

Following the procedure described in the first paragraph of the present section, we first analysed the two elastic-scattering reactions and obtained the solution for the model parameters, as well as the predictions for the em-modified hadronic phase shifts 
derived on their basis \cite{mrw1}. In this paper, we analysed the $\pi^+ p$ and CX databases. By comparing the results of these two PSAs, large effects were found both in two of the model parameters, as well as in the em-modified hadronic phase shifts 
$\tilde{\delta}_{0+}^{1/2}$; significantly smaller differences have been found in two $p$-wave phase shifts.

Assuming the correctness of the absolute normalisation of the bulk of the low-energy $\pi N$ databases, as well as the negligibility of the residual contributions in the em corrections (i.e., of the stage-II effects), these discrepancies can only be 
attributed to the violation of the isospin invariance in the hadronic part of the $\pi N$ interaction at low energies. The effect observed is at the level of $7-8 \%$ in the scattering amplitude below $70$ MeV.

The results of this study agree well with those obtained in the mid 1990s, when the isospin invariance in the $\pi N$ system was first tested by using the then-available experimental information. This agreement is notable given the changes of the databases 
in the meantime (e.g., the CX database has been enlarged by a factor of seven), the analysis methods, and the em corrections applied to the input data. Our result is in disagreement with predictions (for the isospin-breaking effect in the CX 
scattering amplitude) obtained within the framework of the heavy-baryon Chiral-Perturbation Theory, according to which, the expected effects should be around the percent level.

\begin{figure}
\begin{center}
\includegraphics [width=15.5cm] {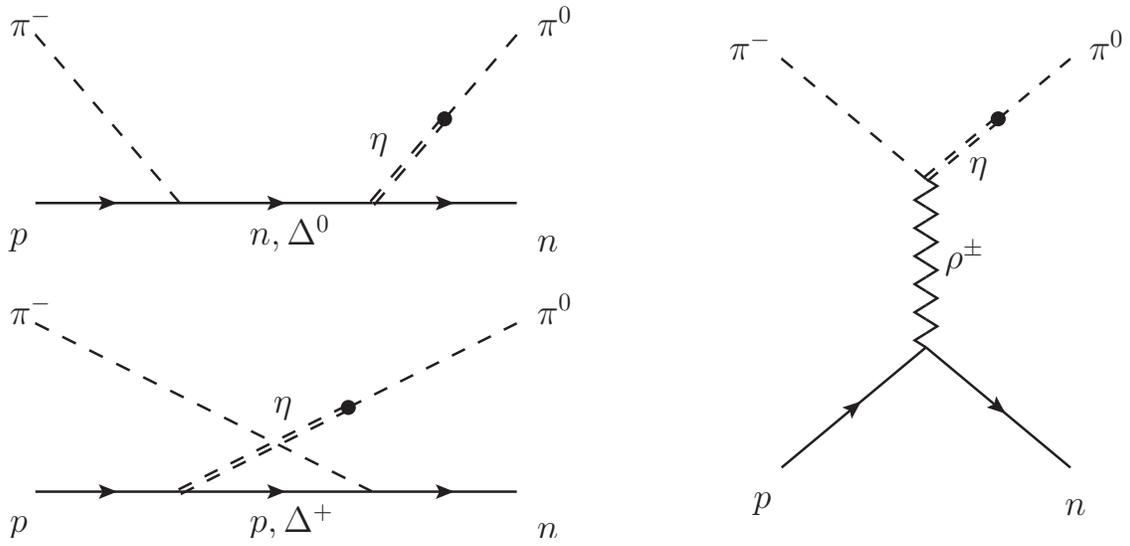}
\caption{\label{fig:IsospinBreakingEtaPi0}Feynman graphs involving the $\eta - \pi^0$ mixing, a potential mechanism for the violation of the isospin invariance in the hadronic part of the $\pi N$ interaction in the case of the 
CX reaction.}
\end{center}
\end{figure}

There are a number of directions which could next be pursued. a) Extensive modifications in our database structure and analysis software should be made in order to include in the fits the AP measurements of Ref.~\cite{meier} and produce a new 
phase-shift solution from the elastic-scattering reactions. Given the goodness of the reproduction of these data with our ZUAS12 prediction (see Section 3.4.5 of Ref.~\cite{mrw1}), only small differences are expected. b) It might be helpful to 
include in the (so-far isospin-invariant) ETH model the isospin-violating Feynman graphs of Fig.~\ref{fig:IsospinBreakingEtaPi0} (as well as the corresponding $\rho^0 - \omega$ graph for elastic scattering), fixing the coupling constants and masses from the 
literature. With the added contributions, one could subsequently investigate whether a significant improvement can be obtained in the description of the $\pi N$ data at low energies. If the graphs of Fig.~\ref{fig:IsospinBreakingEtaPi0} are the dominant ones 
and if the effects observed in the present study are indeed due to the violation of the isospin invariance, then the description of the experimental data (especially when the CX reaction is included in the fits) should improve significantly. c) The 
extrapolation of the amplitudes of the ETH model in the unphysical region, to the Cheng-Dashen point (in order to obtain reliable values of the $\pi N$ $\Sigma$-term), should be seriously investigated.

Finally, it would be interesting to investigate the violation of the isospin invariance using only the results from pionic hydrogen and deuterium (i.e., the data acquired directly at the $\pi N$ threshold) by extracting and comparing the scattering lengths 
and their standard combinations, in a way similar to the one introduced in Ref.~\cite{leisi}. The results of the original experiments (which were performed in the 1990s) are known, as are those for pionic deuterium of the successor experiment; the results 
of this experiment for pionic hydrogen are still marked as preliminary \cite{dg}. It would be interesting to compare all these pieces of information and investigate how they match the picture which is slowly emerging for the $\pi N$ system at low energies. 
The findings of the present work, which corroborate the conclusions of Refs.~\cite{glk,m}, suggest modifications (at least at the low energies) in the established formalism used in the analysis of the $\pi N$ data (e.g., in dispersion-relation schemes). 
Additionally, the extrapolation of the $\pi N$ partial-wave amplitudes in the unphysical region, to the Cheng-Dashen point, must consequently be reconsidered.

\begin{ack}
We acknowledge helpful discussions with W.R. Gibbs. We are grateful to T.P. Gorringe for communicating to us the measurements of Ref.~\cite{jia}. We dedicate this work to W.S. Woolcock (1934-2012) who collaborated with us until the end of 2011. He influenced 
this analysis during its early stages with many important comments and suggestions. Quite unexpectedly, in September 2012, he died after a short illness.

Figure \ref{fig:IsospinBreakingEtaPi0} has been drawn with the software package JaxoDraw \cite{jd}, available from http://jaxodraw.sourceforge.net/.
\end{ack}

\newpage
\begin{table}[h!]
{\bf \caption{\label{tab:CSExp1}}}The low-energy CX experiments in chronological order. The first column contains a label identifying the experiment. Columns $2-5$ contain the number of data points reported in each experiment: DCS stands for the 
differential cross section, LEC for the first three coefficients in the Legendre expansion of the DCS, TCS for the total cross section, and AP for the analysing power. The adjacent column contains the pion laboratory kinetic energy or energy range 
of the experiment. The CM scattering angle or angular interval of the measurements is listed in the last column. The experiment of Ref.~\cite{ss} obtained the scattering length $\tilde{a}^{c0}$ ($=\sqrt{2}b_1$) from a measurement of the width 
of the $1s$ state of pionic hydrogen; therefore, it cannot be placed in any of the categories under Columns $2-5$.
\vspace{0.2cm}
\begin{center}
\begin{tabular}{|l|c|c|c|c|c|c|}
\hline
Experiment & DCS & LEC & TCS & AP & $T$ (MeV) & $\theta$ \\
\hline
BUGG71 \cite{bugg} & & & $1$ & & $90.90$ & \\
DUCLOS73 \cite{duc} & $3$ & & & & $22.60 - 42.60$ & $180^\circ$ \\
SALOMON84 \cite{smpr} & & $6$ & & & $27.40$, $39.30$ & \\
FITZGERALD86 \cite{dhf} & $21$ & & & & $32.48 - 63.21$ & $9.60 - 25.04^\circ$ \\
BAGHERI88 \cite{bagh} & & $12$ & & & $45.60 - 91.70$ & \\
STA{\v S}KO93 \cite{stas} & & & & $4$ & $100.00$ & $75.00 - 130.00^\circ$ \\
FRLE{\v Z}98 \cite{efr} & $6$ & & & & $27.50$ & $4.70 - 50.90^\circ$ \\
GAULARD99 \cite{gaul} & & & & $6$ & $98.10$ & $8.02 - 86.05^\circ$ \\
ISENHOWER99 \cite{ldi} & $40$ & & & & $10.60 - 39.40$ & $9.60 - 168.24^\circ$ \\
SCHROEDER01 \cite{ss} & & & & & $0.00$ & \\
SADLER04 \cite{mesa} & $60$ & & & & $63.86 - 94.57$ & $18.19 - 161.81^\circ$ \\
BREITSCHOPF06 \cite{breit} & & & $9$ & & $38.90 - 96.50$ & \\
JIA08 \cite{jia} & $24$ & & & & $34.37 - 59.68$ & $5.81 - 41.39^\circ$ \\
MEKTEROVI{\' C}09 \cite{mekt} & $140$ & & & & $33.89 - 86.62$ & $18.19 - 161.81^\circ$ \\
\hline
\end{tabular}
\end{center}
\end{table}

\newpage
\begin{table}[h!]
{\bf \caption{\label{tab:OP}}}The list of outliers in the CX database. The rows represent steps in the outlier-identification/elimination process. The columns indicate: the $\chi^2$ value, the number of degrees of freedom in the fit, and the action which 
had to be taken at the specific step. No data was marked for removal at step $6$.
\vspace{0.2cm}
\begin{center}
\begin{tabular}{|c|c|c|l|}
\hline
Step & $\chi^2$ & NDF & Action \\
\hline
$1$ & $400.3$ & $326$ & Freely float FITZGERALD86 at $40.26$ MeV \\
$2$ & $375.1$ & $325$ & Freely float FITZGERALD86 at $36.11$ MeV \\
$3$ & $351.5$ & $324$ & Freely float FITZGERALD86 at $32.48$ MeV \\
$4$ & $333.4$ & $323$ & Exclude BREITSCHOPF06 at $75.10$ MeV \\
$5$ & $324.5$ & $322$ & Freely float FITZGERALD86 at $47.93$ MeV \\
$6$ & $312.8$ & $321$ & \\
\hline
\end{tabular}
\end{center}
\end{table}

\newpage
\begin{table}[h!]
{\bf \caption{\label{tab:DBCX}}}The data sets comprising the truncated database for the CX reaction, the pion laboratory kinetic energy, the number of degrees of freedom for each data set, the scale factor $z_j$ which minimises $\chi_j^2$ (Eq.~(1) of 
Ref.~\cite{mrw1}), the values of $(\chi_j^2)_{min}$, and the p-value associated with the description of each data set. The numbers of this table correspond to the final fit to the data using the $K$-matrix parameterisations (see Section 
\ref{sec:K-Matrix_CX}). In the case of the freely-floated data sets, the scale factor $z_j$ is identical to $\hat{z}_j$ of Eq.(5) of Ref.~\cite{mrw1}.
\vspace{0.2cm}
\begin{center}
\begin{tabular}{|l|c|c|c|c|c|l|}
\hline
Data set & $T$ (MeV)& $(\mathrm{NDF})_j$ & $z_j$ & $(\chi_j^2)_{min}$ & p-value & Comments \\
\hline
BUGG71 & $90.90$ & $1$ & $1.0225$ & $0.1470$ & $0.7014$ & \\
DUCLOS73 & $22.60$ & $1$ & $0.9413$ & $1.2465$ & $0.2642$ & \\
DUCLOS73 & $32.90$ & $1$ & $0.9717$ & $0.2700$ & $0.6033$ & \\
DUCLOS73 & $42.60$ & $1$ & $0.9098$ & $2.3836$ & $0.1226$ & \\
SALOMON84 & $27.40$ & $3$ & $0.9720$ & $2.8685$ & $0.4124$ & \\
SALOMON84 & $39.30$ & $3$ & $0.9937$ & $1.0774$ & $0.7825$ & \\
FITZGERALD86 & $32.48$ & $2$ & $1.5076$ & $2.3635$ & $0.3067$ & freely floated \\
FITZGERALD86 & $36.11$ & $2$ & $1.7103$ & $1.1845$ & $0.5531$ & freely floated \\
FITZGERALD86 & $40.26$ & $2$ & $1.8274$ & $6.1362$ & $0.0465$ & freely floated \\
FITZGERALD86 & $47.93$ & $2$ & $1.4497$ & $1.5402$ & $0.4630$ & freely floated \\
FITZGERALD86 & $51.78$ & $3$ & $1.1236$ & $7.3728$ & $0.0609$ & \\
FITZGERALD86 & $55.58$ & $3$ & $1.0926$ & $2.5611$ & $0.4644$ & \\
FITZGERALD86 & $63.21$ & $3$ & $1.0503$ & $1.2246$ & $0.7471$ & \\
BAGHERI88 & $45.60$ & $3$ & $1.0056$ & $0.1314$ & $0.9878$ & \\
BAGHERI88 & $62.20$ & $3$ & $0.9589$ & $3.4999$ & $0.3208$ & \\
BAGHERI88 & $76.40$ & $3$ & $0.9731$ & $3.2706$ & $0.3518$ & \\
BAGHERI88 & $91.70$ & $3$ & $1.0151$ & $2.8032$ & $0.4230$ & \\
STA{\v S}KO93 & $100.00$ & $4$ & $0.9948$ & $1.4336$ & $0.8383$ & \\
FRLE{\v Z}98 & $27.50$ & $6$ & $1.0902$ & $10.2313$ & $0.1152$ & \\
GAULARD99 & $98.10$ & $6$ & $1.0241$ & $1.1007$ & $0.9815$ & \\
ISENHOWER99 & $10.60$ & $4$ & $1.0203$ & $2.1816$ & $0.7024$ & \\
ISENHOWER99 & $10.60$ & $5$ & $1.0054$ & $1.4611$ & $0.9175$ & \\
ISENHOWER99 & $10.60$ & $6$ & $1.0181$ & $8.0844$ & $0.2320$ & \\
ISENHOWER99 & $20.60$ & $5$ & $0.9803$ & $1.5435$ & $0.9080$ & \\
ISENHOWER99 & $20.60$ & $6$ & $1.0120$ & $8.1813$ & $0.2251$ & \\
\hline
\end{tabular}
\end{center}
\end{table}

\newpage
\begin{table*}
{\bf Table \ref{tab:DBCX} continued}
\vspace{0.2cm}
\begin{center}
\begin{tabular}{|l|c|c|c|c|c|l|}
\hline
Data set & $T$ (MeV)& $(\mathrm{NDF})_j$ & $z_j$ & $(\chi_j^2)_{min}$ & p-value & Comments \\
\hline
ISENHOWER99 & $39.40$ & $4$ & $1.0701$ & $7.1132$ & $0.1300$ & \\
ISENHOWER99 & $39.40$ & $5$ & $1.0597$ & $8.4184$ & $0.1346$ & \\
ISENHOWER99 & $39.40$ & $5$ & $0.9514$ & $5.1617$ & $0.3965$ & \\
SCHROEDER01 & $0.00$ & $1$ & $0.9747$ & $2.5184$ & $0.1125$ & \\
SADLER04 & $63.86$ & $20$ & $0.9548$ & $16.0803$ & $0.7116$ & \\
SADLER04 & $83.49$ & $20$ & $0.9881$ & $11.6506$ & $0.9276$ & \\
SADLER04 & $94.57$ & $20$ & $1.0296$ & $7.2573$ & $0.9958$ & \\
BREITSCHOPF06 & $38.90$ & $1$ & $0.9960$ & $0.1643$ & $0.6852$ & \\
BREITSCHOPF06 & $43.00$ & $1$ & $1.0011$ & $0.0259$ & $0.8721$ & \\
BREITSCHOPF06 & $47.10$ & $1$ & $0.9981$ & $0.0572$ & $0.8110$ & \\
BREITSCHOPF06 & $55.60$ & $1$ & $0.9952$ & $0.2074$ & $0.6488$ & \\
BREITSCHOPF06 & $64.30$ & $1$ & $0.9725$ & $3.7739$ & $0.0521$ & \\
BREITSCHOPF06 & $65.90$ & $1$ & $0.9779$ & $2.3441$ & $0.1258$ & \\
BREITSCHOPF06 & $76.10$ & $1$ & $0.9814$ & $1.6114$ & $0.2043$ & \\
BREITSCHOPF06 & $96.50$ & $1$ & $0.9816$ & $0.6152$ & $0.4328$ & \\
JIA08 & $34.37$ & $4$ & $0.8434$ & $4.9306$ & $0.2945$ & \\
JIA08 & $39.95$ & $4$ & $0.8680$ & $3.1715$ & $0.5295$ & \\
JIA08 & $43.39$ & $4$ & $0.8777$ & $2.5167$ & $0.6416$ & \\
JIA08 & $46.99$ & $4$ & $0.9798$ & $5.1175$ & $0.2754$ & \\
JIA08 & $54.19$ & $4$ & $0.9080$ & $2.0430$ & $0.7279$ & \\
JIA08 & $59.68$ & $4$ & $0.9265$ & $3.2449$ & $0.5177$ & \\
MEKTEROVI{\' C}09 & $33.89$ & $20$ & $1.0239$ & $17.0075$ & $0.6525$ & \\
MEKTEROVI{\' C}09 & $39.38$ & $20$ & $1.0145$ & $14.7514$ & $0.7905$ & \\
MEKTEROVI{\' C}09 & $44.49$ & $20$ & $1.0100$ & $33.1457$ & $0.0325$ & \\
MEKTEROVI{\' C}09 & $51.16$ & $20$ & $1.0357$ & $15.0473$ & $0.7737$ & \\
MEKTEROVI{\' C}09 & $57.41$ & $20$ & $1.0394$ & $19.9034$ & $0.4640$ & \\
MEKTEROVI{\' C}09 & $66.79$ & $20$ & $1.0235$ & $19.4707$ & $0.4914$ & \\
MEKTEROVI{\' C}09 & $86.62$ & $20$ & $1.0019$ & $31.1877$ & $0.0528$ & \\
\hline
\end{tabular}
\end{center}
\end{table*}

\vspace{0.5cm}
\begin{table}
{\bf \caption{\label{tab:ModelParameters}}}The values of the seven parameters of the ETH model obtained from the common fit to the truncated $\pi^+ p$ and CX databases (ZUAS12a solution). The ZUAS12 solution \cite{mrw1}, 
obtained from the PSA of the $\pi^\pm p$ elastic-scattering data for $\mathrm{p}_{min} \approx 1.24 \cdot 10^{-2}$, is shown for comparison.
\vspace{0.2cm}
\begin{center}
\begin{tabular}{|l|c|c|}
\hline
 & This work (ZUAS12a) & ZUAS12 \\
\hline
$G_{\sigma}(GeV^{-2})$ & $30.0 \pm 2.0$ & $27.48 \pm 0.86$ \\
$K_{\sigma}$ & $0.150 \pm 0.058$ & $0.016 \pm 0.034$ \\
$G_{\rho}(GeV^{-2})$ & $59.26 \pm 0.58$ & $54.67 \pm 0.61$ \\
$K_{\rho}$ & $1.65 \pm 0.24$ & $0.66 \pm 0.41$ \\
$g_{\pi N N}$ & $13.43 \pm 0.11$ & $12.84 \pm 0.12$ \\
$g_{\pi N \Delta}$ & $28.97 \pm 0.29$ & $29.77 \pm 0.26$ \\
$Z$ & $-0.353 \pm 0.098$ & $-0.552 \pm 0.056$ \\
\hline
\end{tabular}
\end{center}
\end{table}

\newpage
\begin{table}
{\bf \caption{\label{tab:HessianMatrix}}}The correlation (Hessian) matrix for the seven parameters of the ETH model for the common fit to the truncated $\pi^+ p$ and CX databases.
\vspace{0.2cm}
\begin{center}
\begin{tabular}{|l|c|c|c|c|c|c|c|}
\hline
 & $G_{\sigma}$ & $K_{\sigma}$ & $G_{\rho}$ & $K_{\rho}$ & $g_{\pi N N}$ & $g_{\pi N \Delta}$ &$Z$ \\
\hline
$G_{\sigma}$ & $1.0000$ & $0.4625$ & $0.1180$ & $0.3231$ & $0.3884$ & $-0.5044$ & $-0.2165$ \\
$K_{\sigma}$ & $0.4625$ & $1.0000$ & $0.7696$ & $0.8870$ & $0.8424$ & $-0.9461$ & $0.7475$ \\
$G_{\rho}$ & $0.1180$ & $0.7696$ & $1.0000$ & $0.7081$ & $0.8395$ & $-0.7573$ & $0.7959$ \\
$K_{\rho}$ & $0.3231$ & $0.8870$ & $0.7081$ & $1.0000$ & $0.7794$ & $-0.8712$ & $0.7274$ \\
$g_{\pi N N}$ & $0.3884$ & $0.8424$ & $0.8395$ & $0.7794$ & $1.0000$ & $-0.9053$ & $0.6340$ \\
$g_{\pi N \Delta}$ & $-0.5044$ & $-0.9461$ & $-0.7573$ & $-0.8712$ & $-0.9053$ & $1.0000$ & $-0.6435$ \\
$Z$ & $-0.2165$ & $0.7475$ & $0.7959$ & $0.7274$ & $0.6340$ & $-0.6435$ & $1.0000$ \\
\hline
\end{tabular}
\end{center}
\end{table}
\vspace{0.5cm}

\newpage
\begin{table}
{\bf \caption{\label{tab:ChiSquaresNDF}}}The various $\chi^2$ values obtained in the analysis of the low-energy $\pi N$ data, along with the number of degrees of freedom in each fit, for three values of $\mathrm{p}_{min}$ (the 
confidence level in the statistical tests); these three $\mathrm{p}_{min}$ values correspond to $3$, $2.5$, and $2 \sigma$ effects in the normal distribution. The definitions of the databases are given at the 
end of Section \ref{sec:Method}. Separate fits to the three databases using the ETH model have not been attempted, due to the largeness of the correlations among the model parameters in that case. It must be mentioned that some 
of the $\chi^2$ values of this table, categorised under Ref.~\cite{mrw1}, have not explicitly appeared in that paper.
\vspace{0.2cm}
\begin{center}
\begin{tabular}{|l|c|c|c|c|c|}
\hline
Parametric model & tDB$_+$ \cite{mrw1} & tDB$_-$ \cite{mrw1} & tDB$_0$ (this work) & tDB$_{+/-}$ \cite{mrw1} & tDB$_{+/0}$ (this work) \\
\hline
\multicolumn{6}{|c|}{$\mathrm{p}_{min} \approx 2.70 \cdot 10^{-3}$ ($3 \sigma$)} \\
\hline
$K$-matrix & $434.4/334$ & $397.9/325$ & $333.4/323$ & $825.9/659$ & $765.1/657$ \\
ETH model & $-$ & $-$ & $-$ & $905.4/666$ & $994.4/664$ \\
\hline
\multicolumn{6}{|c|}{$\mathrm{p}_{min} \approx 1.24 \cdot 10^{-2}$ ($2.5 \sigma$)} \\
\hline
$K$-matrix & $427.2/333$ & $371.0/321$ & $312.8/321$ & $792.4/654$ & $737.0/654$ \\
ETH model & $-$ & $-$ & $-$ & $872.9/661$ & $960.5/661$ \\
\hline
\multicolumn{6}{|c|}{$\mathrm{p}_{min} \approx 4.55 \cdot 10^{-2}$ ($2 \sigma$)} \\
\hline
$K$-matrix & $357.0/310$ & $332.3/316$ & $306.4/320$ & $684.4/626$ & $663.3/630$ \\
ETH model & $-$ & $-$ & $-$ & $755.2/633$ & $842.7/637$ \\
\hline
\end{tabular}
\end{center}
\end{table}
\vspace{0.5cm}

\newpage
\begin{table}
{\bf \caption{\label{tab:PhaseShifts}}}The values of the six $s$- and $p$-wave em-modified hadronic phase shifts (in degrees), obtained on the basis of the results of Tables \ref{tab:ModelParameters} (ZUAS12a solution) and 
\ref{tab:HessianMatrix}.
\vspace{0.2cm}
\begin{center}
\begin{tabular}{|c|c|c|c|c|c|c|}
\hline
$T$ (MeV) & $\tilde{\delta}_{0+}^{3/2}$ (S31) & $\tilde{\delta}_{0+}^{1/2}$ (S11) & $\tilde{\delta}_{1+}^{3/2}$ (P33) & $\tilde{\delta}_{1-}^{3/2}$ (P31) & $\tilde{\delta}_{1+}^{1/2}$ (P13) & $\tilde{\delta}_{1-}^{1/2}$ (P11) \\ 
\hline
$20$ & $-2.455(52)$ & $4.535(62)$ & $1.3286(91)$ & $-0.2416(81)$ & $-0.1710(70)$ & $-0.410(12)$ \\
$25$ & $-2.860(54)$ & $5.056(66)$ & $1.884(12)$ & $-0.332(11)$ & $-0.2316(99)$ & $-0.540(17)$ \\
$30$ & $-3.259(56)$ & $5.519(70)$ & $2.515(15)$ & $-0.430(15)$ & $-0.295(13)$ & $-0.670(23)$ \\
$35$ & $-3.657(57)$ & $5.937(72)$ & $3.222(18)$ & $-0.534(19)$ & $-0.360(17)$ & $-0.795(29)$ \\
$40$ & $-4.055(58)$ & $6.318(75)$ & $4.007(20)$ & $-0.643(24)$ & $-0.427(21)$ & $-0.914(35)$ \\
$45$ & $-4.455(59)$ & $6.666(78)$ & $4.873(22)$ & $-0.756(29)$ & $-0.495(25)$ & $-1.023(42)$ \\
$50$ & $-4.858(61)$ & $6.987(82)$ & $5.824(24)$ & $-0.874(34)$ & $-0.562(29)$ & $-1.123(49)$ \\
$55$ & $-5.264(64)$ & $7.282(87)$ & $6.865(26)$ & $-0.994(40)$ & $-0.630(34)$ & $-1.211(57)$ \\
$60$ & $-5.673(69)$ & $7.554(93)$ & $8.000(28)$ & $-1.118(46)$ & $-0.697(39)$ & $-1.287(65)$ \\
$65$ & $-6.087(75)$ & $7.80(10)$ & $9.238(30)$ & $-1.245(52)$ & $-0.763(44)$ & $-1.350(74)$ \\
$70$ & $-6.504(82)$ & $8.04(11)$ & $10.586(34)$ & $-1.374(59)$ & $-0.828(50)$ & $-1.398(83)$ \\
$75$ & $-6.924(92)$ & $8.25(12)$ & $12.053(39)$ & $-1.506(67)$ & $-0.893(56)$ & $-1.431(92)$ \\
$80$ & $-7.35(10)$ & $8.44(13)$ & $13.648(47)$ & $-1.640(74)$ & $-0.956(62)$ & $-1.45(10)$ \\
$85$ & $-7.78(11)$ & $8.62(14)$ & $15.383(57)$ & $-1.775(82)$ & $-1.017(68)$ & $-1.45(11)$ \\
$90$ & $-8.21(13)$ & $8.78(16)$ & $17.268(70)$ & $-1.912(90)$ & $-1.078(75)$ & $-1.43(12)$ \\
$95$ & $-8.64(14)$ & $8.93(17)$ & $19.318(85)$ & $-2.051(99)$ & $-1.136(82)$ & $-1.40(14)$ \\
$100$ & $-9.08(16)$ & $9.06(19)$ & $21.54(10)$ & $-2.19(11)$ & $-1.193(89)$ & $-1.35(15)$ \\
\hline
\end{tabular}
\end{center}
\end{table}

\newpage
\begin{table}
{\bf \caption{\label{tab:ReproductionOfPTCS}}}Reproduction of the $\pi^- p$ PTCSs which had not been used in Ref.~\cite{mrw1}. The first four columns correspond to a label identifying the measurement, the pion laboratory 
kinetic energy, the laboratory-angle cut ($\theta_L$), and the PTCS measurement. The three adjacent columns contain the predictions obtained on the basis of our two PSAs, namely of Ref.~\cite{mrw1} and of the present work. The column 
marked as `$\sigma_{EL}$' contains the prediction for the $\pi^- p$ elastic-scattering PTCS (for the particular $\theta_L$ value) obtained from the ZUAS12 solution of Ref.~\cite{mrw1}. The next column contains the prediction for the 
CX TCS also obtained from the ZUAS12 solution of Ref.~\cite{mrw1}. The last column contains the prediction for the CX TCS obtained from the solution of the present work.
\vspace{0.2cm}
\begin{center}
\begin{tabular}{|l|c|c|c|c|c|c|}
\hline
Data point & $T$ (MeV) & $\theta_L$ (deg) & $\sigma$ (mb) & $\sigma_{EL}$ (mb) \cite{mrw1} & $\sigma_{CX}$ (mb) \cite{mrw1} & $\sigma_{CX}$ (mb) \\ 
\hline
FRIEDMAN90 \cite{frn} & $50.00$ & $30.00$ & $8.5 \pm 0.6$ & $2.15 \pm 0.13$ & $6.089 \pm 0.048$ & $6.618 \pm 0.050$ \\ 
KRISS97 \cite{bjk} & $80.00$ & $30.00$ & $14.6 \pm 0.6$ & $2.958 \pm 0.076$ & $11.290 \pm 0.074$ & $11.808 \pm 0.089$ \\ 
KRISS97 \cite{bjk} & $99.20$ & $30.00$ & $23.4 \pm 1.1$ & $4.594 \pm 0.040$ & $17.34 \pm 0.15$ & $17.72 \pm 0.17$ \\
\hline
\end{tabular}
\end{center}
\end{table}

\newpage
\begin{table}[h!]
{\bf \caption{\label{tab:CSExp2}}}The scale factors $\hat{z}_j$ of the CX data sets, which are appropriate for testing their absolute normalisation relative to the ZUAS12 predictions \cite{mrw1}, listed separately for the differential 
cross sections (upper part), total cross sections (second part), analysing powers (third part), and the results for the coefficients of the Legendre expansion of the differential cross section (last part). The four FITZGERALD86 data sets 
which had been floated freely in Section \ref{sec:K-Matrix_CX} have not been included. Not listed in the table is also the result of Ref.~\cite{ss} for the isovector scattering length $b_1$. The quantity $\Delta \hat{z}_j$ is the total 
uncertainty (see end of Section 2.2 of Ref.~\cite{mrw1}).
\vspace{0.2cm}
\begin{center}
\begin{tabular}{|l|c|c|c|}
\hline
Data set & $T$ (MeV) & $\hat{z}_j$ & $\Delta \hat{z}_j$ \\
\hline
DUCLOS73 & $22.60$ & $1.03$ & $0.14$ \\
DUCLOS73 & $32.90$ & $1.09$ & $0.13$ \\
DUCLOS73 & $42.60$ & $0.95$ & $0.12$ \\
FITZGERALD86 & $51.78$ & $1.29$ & $0.10$ \\
FITZGERALD86 & $55.58$ & $1.250$ & $0.099$ \\
FITZGERALD86 & $63.21$ & $1.203$ & $0.093$ \\
FRLE{\v Z}98 & $27.50$ & $1.431$ & $0.099$ \\
ISENHOWER99 & $10.60$ & $1.45$ & $0.12$ \\
ISENHOWER99 & $10.60$ & $1.331$ & $0.081$ \\
ISENHOWER99 & $10.60$ & $1.307$ & $0.057$ \\
ISENHOWER99 & $20.60$ & $1.205$ & $0.052$ \\
ISENHOWER99 & $20.60$ & $1.229$ & $0.047$ \\
ISENHOWER99 & $39.40$ & $1.48$ & $0.11$ \\
ISENHOWER99 & $39.40$ & $1.245$ & $0.045$ \\
ISENHOWER99 & $39.40$ & $1.089$ & $0.042$ \\
SADLER04 & $63.86$ & $1.050$ & $0.068$ \\
SADLER04 & $83.49$ & $1.045$ & $0.053$ \\
SADLER04 & $94.57$ & $1.063$ & $0.047$ \\
JIA08 & $34.37$ & $1.04$ & $0.12$ \\
JIA08 & $39.95$ & $1.00$ & $0.12$ \\
JIA08 & $43.39$ & $0.94$ & $0.13$ \\
JIA08 & $46.99$ & $1.09$ & $0.14$ \\
JIA08 & $54.19$ & $0.94$ & $0.13$ \\
JIA08 & $59.68$ & $0.99$ & $0.12$ \\
\hline
\end{tabular}
\end{center}
\end{table}

\newpage
\begin{table*}
{\bf Table \ref{tab:CSExp2} continued}
\vspace{0.2cm}
\begin{center}
\begin{tabular}{|l|c|c|c|}
\hline
Data set & $T$ (MeV) & $\hat{z}_j$ & $\Delta \hat{z}_j$ \\
\hline
MEKTEROVI{\' C}09 & $33.89$ & $1.213$ & $0.040$ \\
MEKTEROVI{\' C}09 & $39.38$ & $1.182$ & $0.032$ \\
MEKTEROVI{\' C}09 & $44.49$ & $1.160$ & $0.031$ \\
MEKTEROVI{\' C}09 & $51.16$ & $1.181$ & $0.033$ \\
MEKTEROVI{\' C}09 & $57.41$ & $1.167$ & $0.031$ \\
MEKTEROVI{\' C}09 & $66.79$ & $1.127$ & $0.032$ \\
MEKTEROVI{\' C}09 & $86.62$ & $1.056$ & $0.030$ \\
\hline
BUGG71 & $90.90$ & $1.068$ & $0.061$ \\
BREITSCHOPF06 & $38.90$ & $1.12$ & $0.10$ \\
BREITSCHOPF06 & $43.00$ & $1.17$ & $0.15$ \\
BREITSCHOPF06 & $47.10$ & $1.11$ & $0.12$ \\
BREITSCHOPF06 & $55.60$ & $1.077$ & $0.093$ \\
BREITSCHOPF06 & $64.30$ & $0.967$ & $0.069$ \\
BREITSCHOPF06 & $65.90$ & $0.995$ & $0.067$ \\
BREITSCHOPF06 & $76.10$ & $0.994$ & $0.065$ \\
BREITSCHOPF06 & $96.50$ & $0.999$ & $0.039$ \\
\hline
STA{\v S}KO93 & $100.00$ & $0.91$ & $0.11$ \\
GAULARD99 & $98.10$ & $0.962$ & $0.058$ \\
\hline
SALOMON84 & $27.40$ & $1.098$ & $0.056$ \\
SALOMON84 & $39.30$ & $1.131$ & $0.059$ \\
BAGHERI88 & $45.60$ & $1.149$ & $0.036$ \\
BAGHERI88 & $62.20$ & $1.042$ & $0.039$ \\
BAGHERI88 & $76.40$ & $1.038$ & $0.036$ \\
BAGHERI88 & $91.70$ & $1.063$ & $0.039$ \\
\hline
\end{tabular}
\end{center}
\end{table*}

\clearpage

\clearpage

\clearpage

\clearpage

\clearpage

\clearpage

\clearpage

\clearpage
\clearpage

\end{document}